\documentclass[fleqn,usenatbib,useAMS]{mnras}

\usepackage{newtxtext,newtxmath}

\usepackage[T1]{fontenc}
\usepackage{ae,aecompl}

\usepackage{graphicx}
\usepackage{amsmath}
\usepackage{amssymb}

\usepackage[usenames,dvipsnames]{color}
\usepackage{enumerate}

\usepackage{fixltx2e}
\usepackage{amsmath}
\usepackage{float}
\usepackage[caption = false]{subfig}
\usepackage{multirow}
\include{journals}

\newcommand{\tension}{G\mu/c^2}
\newcommand{\thetastr}{\theta_{\rm str}}
\newcommand{\Rgeo}{R_{\rm \,geo}}
\newcommand{\Rmicro}{R_{\rm \, micro}}

\newcommand{\Gammalens}{\Tilde{\Gamma}}
\newcommand{\Rlens}{R_{\rm \,lens}}
\newcommand{\Plens}{P_{\rm \,lens}}
\newcommand{\Nedge}{N_{\rm \,edge}}

\newcommand{\nlens}{n_{\rm \,lens }}
\newcommand{\nedge}{n_{\rm \,edge }}
\newcommand{\ncl}{n_{\rm \,CL }}
\newcommand{\nmax}{n_{\rm \,max }}

\newcommand{\epslens}{\epsilon_{\rm \, lens}}
\newcommand{\epscl}{\epsilon_{\rm \, CL}}
\newcommand{\epsedge}{\epsilon_{\rm \, edge}}

\newcommand{\dma}{d_{\rm max}}
\newcommand{\dmi}{d_{\rm min}}
\newcommand{\delma}{\delta_{\rm max}}
\newcommand{\delmi}{\delta_{\rm min}}

\newcommand{\Nobs}{N_{\rm obs}}
\newcommand{\Nav}{\bar N}
\newcommand{\Nrep}{N_{\rm rep}}
\newcommand{\nav}{\bar n}

\newcommand{\vcom}{v_{\rm com}}
\newcommand{\vcomperp}{v_{\rm com, \, \perp}}
\newcommand{\vosc}{v_{\rm osc}}
\newcommand{\vinter}{v_{\rm i}}

\newcommand{\Tobs}{T_{\rm obs}}
\newcommand{\tpass}{t_{\rm pass}}
\newcommand{\tobs}{t_{\rm obs}}
\newcommand{\tlens}{t_{\rm lens}}
\newcommand{\tlensm}{t_{\rm lens,\, max}}
\newcommand{\tosc}{t_{\rm osc}}

\newcommand{\rstring}{d}
\newcommand{\rsource}{D}
\newcommand{\rhodm}{\rho_{\rm DM}}
\newcommand{\Omdm}{\Omega_{\rm DM}}
\newcommand{\rhocrit}{\rho_{\rm crit}}

\newcommand{\dd}{{\rm d}}
\newcommand{\dndlogl}{\left( \frac{\dd \, n}{\dd \, {\rm ln}\, l} \right)}
\newcommand{\dndl}{\left( \frac{\dd \, n}{\dd \, l} \right)}
\newcommand{\dndlinline}{(\dd \, n/\dd \, l)}
\newcommand{\dndloglinline}{(\dd \, n/\dd \, {\rm ln}\, l)}

\newcommand{\lp}{\emph{left panel}}
\newcommand{\Lp}{\emph{Left panel}}
\newcommand{\rp}{\emph{right panel}}
\newcommand{\Rp}{\emph{Right panel}}

\newcommand{\lta}{{\mathbin{\lower 3pt\hbox
   {$\rlap{\raise 5pt\hbox{$\char'074$}}\mathchar"7218$}}}}
\newcommand{\gta}{{\mathbin{\lower 3pt\hbox
   {$\rlap{\raise 5pt\hbox{$\char'076$}}\mathchar"7218$}}}}
\usepackage{hyperref}
\hypersetup{
    colorlinks=true,
    linkcolor=blue,
    citecolor=blue,
    filecolor=black,
    urlcolor=blue,
}

\title[Cosmic Superstring Detection]{Prospects of Cosmic Superstring Detection through Microlensing of Extragalactic Point-Like Sources} 

\author[Chernoff, Goobar \& Renk]{David F. Chernoff\thanks{chernoff@astro.cornell.edu},$^1$ Ariel Goobar,$^2$ Janina J.~Renk\thanks{janina.renk@fysik.su.se}$^2$ \\
$^{1}$Department of Astronomy, Cornell University, Ithaca, NY 14853, USA\\
$^{2}$The Oskar Klein Centre for Cosmoparticle Physics, Department of Physics, Stockholm University\\
        AlbaNova University Centre, SE-106 91 Stockholm, Sweden\\
}

\date{Accepted XXX. Received YYY; in original form ZZZ}

\pubyear{2019}

\begin{document}
\label{firstpage}
\pagerange{\pageref{firstpage}--\pageref{lastpage}}
\maketitle

\begin{abstract}
The existence of cosmic superstrings may be probed by astronomical time domain surveys.  When crossing the line of sight to point-like sources, strings produce a distinctive microlensing signature.  We consider two avenues to hunt for a relic population of superstring loops: frequent monitoring of (1) stars in Andromeda, lensed by loops in the haloes of the Milky-Way and Andromeda and (2) supernovae at cosmological distances, lensed by loops in the intergalactic medium. We assess the potential of such experiments to detect and/or constrain strings with a range of tensions, $10^{-15} \lesssim G \mu/c^2 \lesssim 10^{-6}$. The practical sensitivity is tied to cadence of observations which we explore in detail.  We forecast that high-cadence monitoring of $\sim 10^5$ stars on the far side of Andromeda over a year-long period will detect microlensing events if $G\mu/c^2 \sim 10^{-13}$, while $\sim 10^6$ stars will detect events if $10^{-13.5} <G\mu/c^2 < 10^{-11.5}$; the upper and lower bounds of the accessible tension range continue to expand as the number of stars rises. We also analyse the ability to reject models in the absence of fluctuations. While challenging, these studies are within reach of forthcoming time-domain surveys. Supernova observations can hypothetically constrain models with $10^{-12} < G\mu/c^2 < 10^{-6}$ without any optimization of the survey cadence. However, the event rate forecast suggests it will be difficult to reject models of interest. As a demonstration, we use observations from the {\it Pantheon} Type Ia supernova cosmology data-set to place modest constraints on the number density of cosmic superstrings in a poorly tested region of the parameter space.

\end{abstract}




\section{Introduction}
We explore the expected rate of lensing of background optical sources induced gravitationally by string-like objects \citep{Vilenkin:1981zs}
which naturally arise in two prominent theoretical scenarios in fundamental physics. {\it Cosmic strings} \citep{Kibble:1976sj} are one-dimensional topological defects formed by spontaneous symmetry breaking in the early universe; {\it cosmic superstrings} \citep{Witten:1985fp} are stable elements of the string theory framework stretched to macroscopic size. General relativistic light-bending by both types is similar, since they are effectively one-dimensional sources with analogous stress energy tensors in 3+1 dimensional space time. For reviews of cosmic strings see \cite{Vilenkin:2000jqa} and of superstrings \cite{Chernoff:2014cba}.

We consider two observational scenarios: (1) lensing of stars in the nearest major galaxy, Andromeda (M31), by strings in the Milky-Way and Andromeda haloes and (2) lensing of supernovae at cosmological distance by strings in the intergalactic medium (IGM). In both situations, as light passes close to a segment of string,  two images of the background source form for a geometry with suitably aligned source, lens and observer. The image separation is too small to be spatially resolved for string tensions of current interest so that only the modified total flux is potentially observable. This is string {\em microlensing}. It imprints a distinct signal: the flux of the source is enhanced by a fixed factor of two during lensing. This tell-tale signature makes microlensing events caused by strings easily distinguishable from astrophysical sources, or other exotic lenses like primordial black holes or axion miniclusters. We estimate the detection rate of these events in observational campaigns to assess whether limits can already be set with present data. Further we address how these bounds can be tightened in the future.

Why strings? Early interest in Grand Unified Theory (GUT) cosmic strings related to the possibility that strings might actively create the density fluctuations that subsequently evolved into structures in the present day universe. However, the observation of acoustic peaks of the cosmic microwave background (CMB) power spectrum ruled out cosmic strings as the primary source of fluctuations and, moreover, strongly supported the occurrence of an inflationary period \citep{Spergel:2006hy}. 

Today, string theory provides a consistent framework for the unification of all of nature's known force fields and matter content. That framework provides multiple, promising routes to realize inflation \citep[see][for a review]{Baumann:2014nda}, many leading to the creation of macroscopic, effectively one-dimensional objects at inflation's end. These are nothing more than the basic constituents of the theory stretched to huge sizes by the rapid expansion of the universe. Superstrings behave in ways that are similar to cosmic strings: they form a network of long, horizon-crossing segments and sub-horizon loops. For current observationally allowed tensions, the loops are the dominant component of interest. In this paper we consider the prospects for a direct search for {\em superstring loops}, fossil remnants of the early universe. 

The tension $\mu$ is the primary parameter that controls the physical properties of the network (e.g. the cosmological fraction of loops and long horizon-crossing segments, the present-day characteristic size and number density of loops, etc.). In the original studies of GUT strings, the GUT energy scale, $\Lambda \sim 10^{16}$ GeV, set the dimensionless string tension $G \mu/c^2 \sim (\Lambda/E_p)^2 \sim 10^{-6}$, where $G$ is Newton's constant, $c$ is the speed of light and $E_p$ is the Planck energy. In string theory scenarios, the observed string tension is not generally fixed at a particular scale like the string scale. In warped geometries, for example, the energy scale for strings is exponentially smaller.

Empirical upper bounds on $G \mu/c^2$ have been derived from null results for experiments involving gravitational lensing \citep{Vilenkin:1981zs,Hogan:1984unknown,Vilenkin:1984ea,deLaix:1997dj,Bernardeau:2000xu,Sazhin:2003cp,Sazhin:2006fe,Christiansen:2008vi}, gravitational wave background and bursts \citep{Vachaspati:1984gt,Economou:1991bc,Battye:1997ji,Damour:2000wa,Damour:2001bk,Damour:2004kw,Siemens:2006vk,Hogan:2006we,Siemens:2006yp,Abbott:2006vg,Abbott:2009rr,Abbott:2009ws,Aasi:2013vna,TheLIGOScientific:2016dpb}, pulsar timing \citep{Bouchet:1989ck,Caldwell:1991jj,Kaspi:1994hp,Jenet:2006sv,DePies:2007bm,Blanco-Pillado:2017oxo,Blanco-Pillado:2017rnf} and CMB radiation \citep{Smoot:1992td,Bennett:1996ce,Pogosian:2003mz,Pogosian:2004ny,Tye:2005fn,Wyman:2005tu,Pogosian:2006hg,Seljak:2006bg,Spergel:2006hy,Bevis:2007qz,Fraisse:2006xc,Pogosian:2008am,Ade:2013xla}.

\cite{Pshirkov:2009vb} presented the first observational constraints on the density of strings based in part on the physics of microlensing. They utilized X-ray data for Sco X-1 ({\it RXTE}) and precision optical photometry ({\it CoRoT}) towards 30 local Galactic sources plus pulsar timing. They treated the string loop density and tension as independent parameters and inferred constraints on the mean string loop density $\Omega_{loop} \lta 10^{-3}$ at $\tension \sim 10^{-14}$ from the lack of flux and timing variations. \cite{Tuntsov:2010fu} quantified the microlensing-induced variation in {\it SDSS} quasars and limited the total density in loops plus long, horizon-crossing strings to be $\Omega \lta 10^{-2}$ at tensions $10^{-13} \lta \tension \lta 10^{-11}$.

All such bounds are model-dependent with respect to secondary parameters \citep{Chernoff:2014cba}. Even when these are fixed a variety of observational and astrophysical uncertainties remain.  CMB power spectrum fits rely on well-established gross properties of large-scale string networks and are relatively secure. Limits from optical lensing in fields of background galaxies rely on the theoretically well-understood deficit angle geometry of a string in space time but require a precise understanding of observational selection effects.  Limits from big bang nucleosynthesis (BBN; see \citealt{Abbott:2017mem} for discussion of the BBN limit) rely on changes to the expansion rate
from extra gravitational energy density but only constrain the strings formed prior to that epoch. Roughly speaking, the combination of approaches implies $G \mu/c^2 \, \lta \, 3 \times 10^{-8} - 3 \times 10^{-7}$ \cite[for comparisons of limits]{Siemens:2006yp,Abbott:2017mem}. More stringent results generally invoke additional assumptions \citep{Battye:2010xz}. Gravitational wave experiments \citep{Siemens:2006yp,Abbott:2009ws} and long-term pulsar timing \citep{Jenet:2006sv} are sensitive to the unknown number of radiating cusps on each loop and to the number and size of newly formed loops. The strongest bound $G \mu/c^2 \, \lta \, 1.5 \times 10^{-11}$ follows from the lack of detection of the stochastic background by pulsar timing experiments \citep{Blanco-Pillado:2017oxo,Blanco-Pillado:2017rnf}. It depends upon network simulations of the loop distribution and specific assumptions about how gravitational backreaction modifies the shape of evaporating loops. In the future LISA-like experiments may achieve limits as low as $G \mu/c^2 \, \sim 5.8 \, \times 10^{-18}$ \citep{Blanco-Pillado:2017rnf}.

In short, cosmic superstrings are produced during inflation in well-studied theoretical models and observations imply tensions substantially lower than the original GUT-inspired strings. New observations can play an important role constraining $G \mu/c^2$ from above. Multiple, overlapping approaches are needed to minimize model-dependent aspects.  There is no known theoretical impediment to the magnitude of $G \mu/c^2$ being either comparable to or much lower than the current observational upper limits. Microlensing searches for strings are sensitive to the presence of strings in the local (Galactic and/or low-z) Universe.

This article is organized as follows: we briefly review the cosmology of cosmic superstrings in \autoref{sec:cosmo} and describe our modelling of the Dark Matter (DM) and string loop distributions. \hyperref[sec:ML]{Section~\ref*{sec:ML}} discusses microlensing by strings and its unique signature;  \autoref{sec:ObsLimits} covers various experimental limitations on observing that signature. The calculation of the time-scales of the lensing events of stars in M31 and distant sources lensed by strings in the IGM is given in \autoref{sec:time-scales}. \hyperref[sec:simTech]{Section~\ref*{sec:simTech}} explains the method for simulating the expected number of lensing events for a given target and survey; the results of these calculations are discussed in \autoref{sec:Results}, where we summarise our main results in \autoref{fig:CombinedRates}. We conclude with a summary in \autoref{sec:concl}.

%
\section{Cosmology of Superstrings} \label{sec:cosmo}
%

Today's superstrings are descendants of horizon-crossing strings created at the end of the inflationary era.
\citeauthor{Chernoff:2017fll} (\citeyear{Chernoff:2017fll}, hereafter CT18) review brane inflation and the resultant superstring loop population. Brane inflation is a well-studied scenario in which the collision of 3-branes in a warped throat attached to the bulk space heats the universe, creates the big bang and produces a network of superstrings by a Kibble-like mechanism. Realistic compactifications may harbor many throats, each containing its own set of strings. There are many uncertainties regarding the number of species of strings and the reduced efficiency of intercommutation, the process of breaking and rejoining string elements. These considerations generally lead to larger number densities of superstrings than predicted for a single species of field theory strings at the GUT scale. CT18 gathered together all the differences between superstrings and a single string species having unit intercommutation probability (hereafter, SSSUIP) into a single factor ${\cal  G}$, estimated a theoretical range $1 \le {\cal G} < 10^4$ and took ${\cal G}=10^2$ as a best estimate. Here, we will regard $\tension$ and ${\cal G}$ as the key, unknown parameters of string theory. In our explorations we will allow tension to vary freely below current observational upper limits and use ${\cal G}=10^2$ in our numerical work. We will forecast limits on both parameters that can be placed by microlensing experiments.

Simulations of string networks show that breaking/rejoining collisions of long horizon-crossing string segments generate loops. Further, this loop size is tied to the scale of the horizon at the time of formation. Small loops are formed at early times, larger ones at later times. Loop densities dilute as the universe expands. If loops were stable then the smallest loops would dominate both the total number and total length of strings in the universe today.\footnote{They would eventually create a monopole-like problem and overclose the universe.} In fact, loops oscillate, emit gravitational radiation and eventually disappear. The lifetime is $t_{\rm life} \sim (l/c)/(\Gamma G \mu/c^2)$ where $l$ is the loop length and $\Gamma \sim 50$ is a loop-dependent dimensionless number. Roughly speaking, the smallest surviving loops have the highest number density today.  Large loops are born moving quickly, $\vinter \sim 0.1-0.3 \,c$ \citep{Bennett:1989ch} and slow down as the universe expands.

One reason that string networks have been of cosmological interest is because the dissipative processes of string collision and gravitational radiation robustly establish a scaling solution in which long strings, loops and remnant radiation supply fixed fractions of the critical density. Models with a string network track traditional cosmological models but also include small amounts of the string-related components.

Both superstring and GUT string networks behave in this manner but the cosmology of superstrings is distinct because $G \mu/c^2$ is much smaller than the GUT value. The logic is the following:
\begin{enumerate}[1)]
\item Loops born with $l < l_g \equiv t_H \Gamma G \mu/c$ will have evaporated by today, where $t_H$ is the characteristic age of the Universe;
\item Today, the preponderance of loops in the universe are those born slightly (factor of 2) larger than $l_g$;
\item Parametrically, lowering $\mu$ implies smaller loops formed when the horizon was smaller and when the universe was older; the number densities of such loops today are larger;
\item Formation at larger $z$ implies that the initial loop peculiar centre of mass motion suffers more damping on account of the universe's expansion.
\end{enumerate}

To summarize: intercommutations of the cosmological string network
chop out loops having mildly relativistic velocities $\vinter$.  Once
an individual loop moves freely through space (clear of the network,
other loops and not self-intersecting) its peculiar motion damps as
the Universe expands. The length of the loop diminishes by emission of
gravitational radiation and it has a lifetime inversely proportional
to string tension. An old, low tension slowly moving string loop may
be captured into a growing galactic potential if it is in the right
place at the right time; or, it can linger for its entire lifetime in
intergalactic space. In either scenario the loop's typical asymmetric
emission generates a recoil that accelerates the loop's center of
mass, the so-called rocket effect.\footnote{The degree of asymmetry
  depends upon the details of its oscillation.  When a cusp is present
  the average radiated momentum is $\sim 5-10$\% of the total radiated
  energy.}  The acceleration given the loop grows secularly since the
loop mass decreases while the rate of emission is roughly constant.

If the direction of the rocket impulse is fixed then an intergalactic
loop smoothly reaccelerates to high velocities. There are two
complications \citep{Chernoff:2009tp}.  (1) A loop accreted by
structure before the rocket effect becomes significant is confined to
the galactic potential, the rocket's force is averaged over a periodic
orbit and the net work done on the center of mass is adiabatically
suppressed. Eventually the force grows large and dislodges the loop
from the potential and it returns to the IGM. The escape condition is
that the rocket acceleration exceed that of gravitational binding. (2)
As the loop shrinks the direction of the rocket is expected to evolve
so that the velocity may not grow monotonically.

There are two microlensing populations of interest to us: the loops
confined to galactic potentials (such as the Galaxy and M31) and those
moving freely in intergalactic space (along a typical IGM line of
sight to a supernova). The differential number density of loops per
invariant length $dn/d\ell$ rises at $\ell \to 0$. Experimental
forecasts weight the number density by the rate or cross section per
loop, typically, $\propto \ell$. The peak of the weighted number
density is $\ell \sim (2/3)\ell_g$.  In general, the smallest loops
with size $\ell \gta \ell_g$ dominate the observational forecasts for
a given population.

The velocity of an IGM loop with $\ell=\ell_g$ today reaches $\sim
0.04$c when emission has 10\% asymmetry and fixed direction.  On the
other hand, loops that the Galaxy accreted during formation have much
lower halo velocities. The criterion for ejection is $\ell/\ell_g \lta
0.5-3$ for orbital radii $3-20$ kpc respectively.  Larger loops
remain bound with typical halo velocities.  The observational possibilities for
both IGM and Galactic populations are dominated by the loops with
$\ell \sim \ell_g$.  The characteristic halo velocity of those bound to the
Galaxy is $\sim 220$ km/s.  However, the
velocity of those in the IGM depends upon the rate of wander in the
rocket direction compared to the rate of evaporation.  If the wander
is slow the speed estimate $0.04$c or $12,000$ km/s; if the wander is fast
then the motions cannot be less than the typical IGM mass element,
$\gta 600$ km/s. We will test both of these limits for the IGM loops.

%
\begin{figure*}
\begin{minipage}{0.45\linewidth}
     \centering
     \includegraphics[trim= 0 0 0 0 0, clip, width=7.9cm]{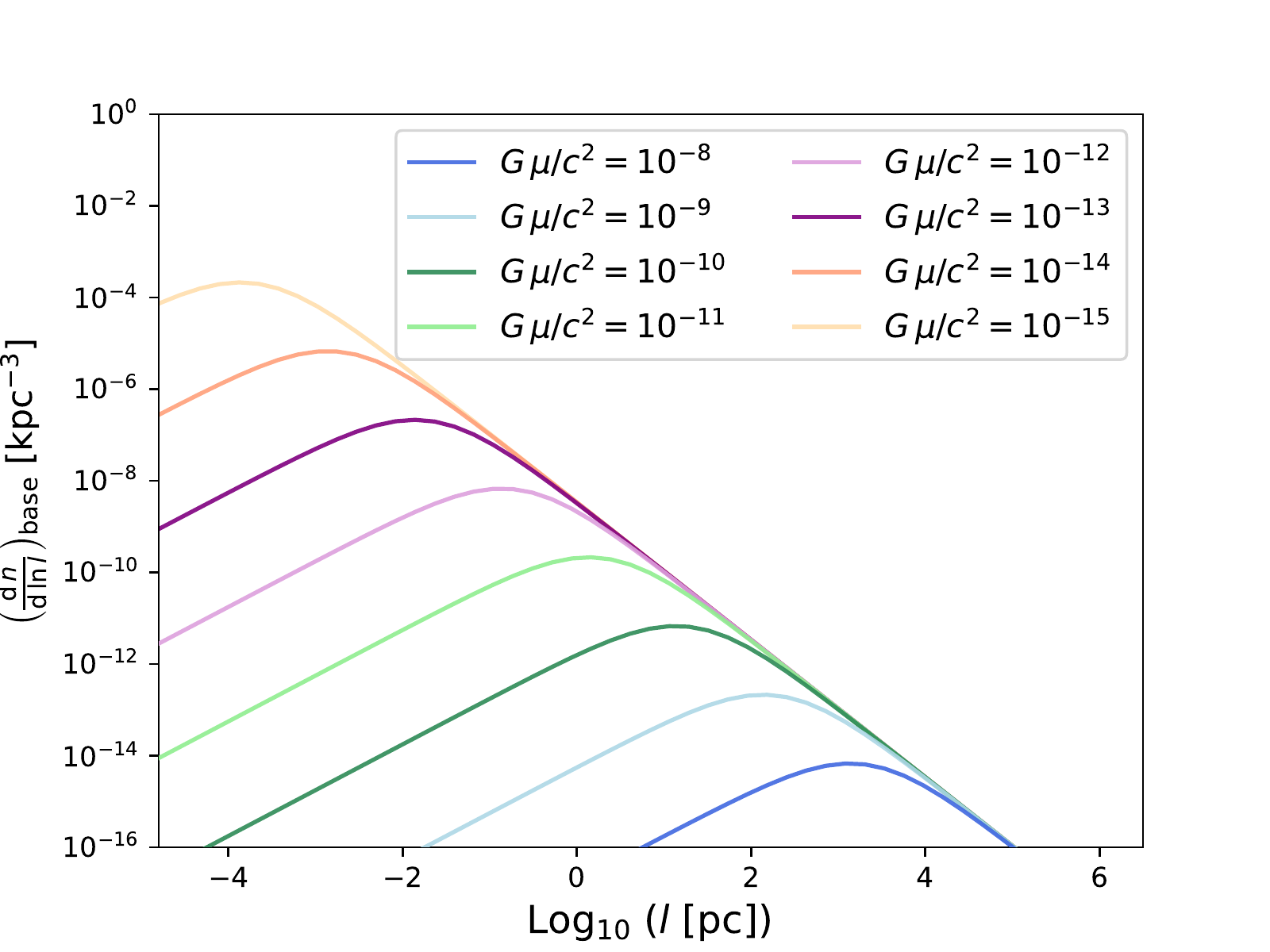}
     \vspace{0.05cm}
    \end{minipage}
    \begin{minipage}{0.45\linewidth}
      \centering\includegraphics[trim= 0.0 0.0 0 0.0 0, clip, width=7.9cm]{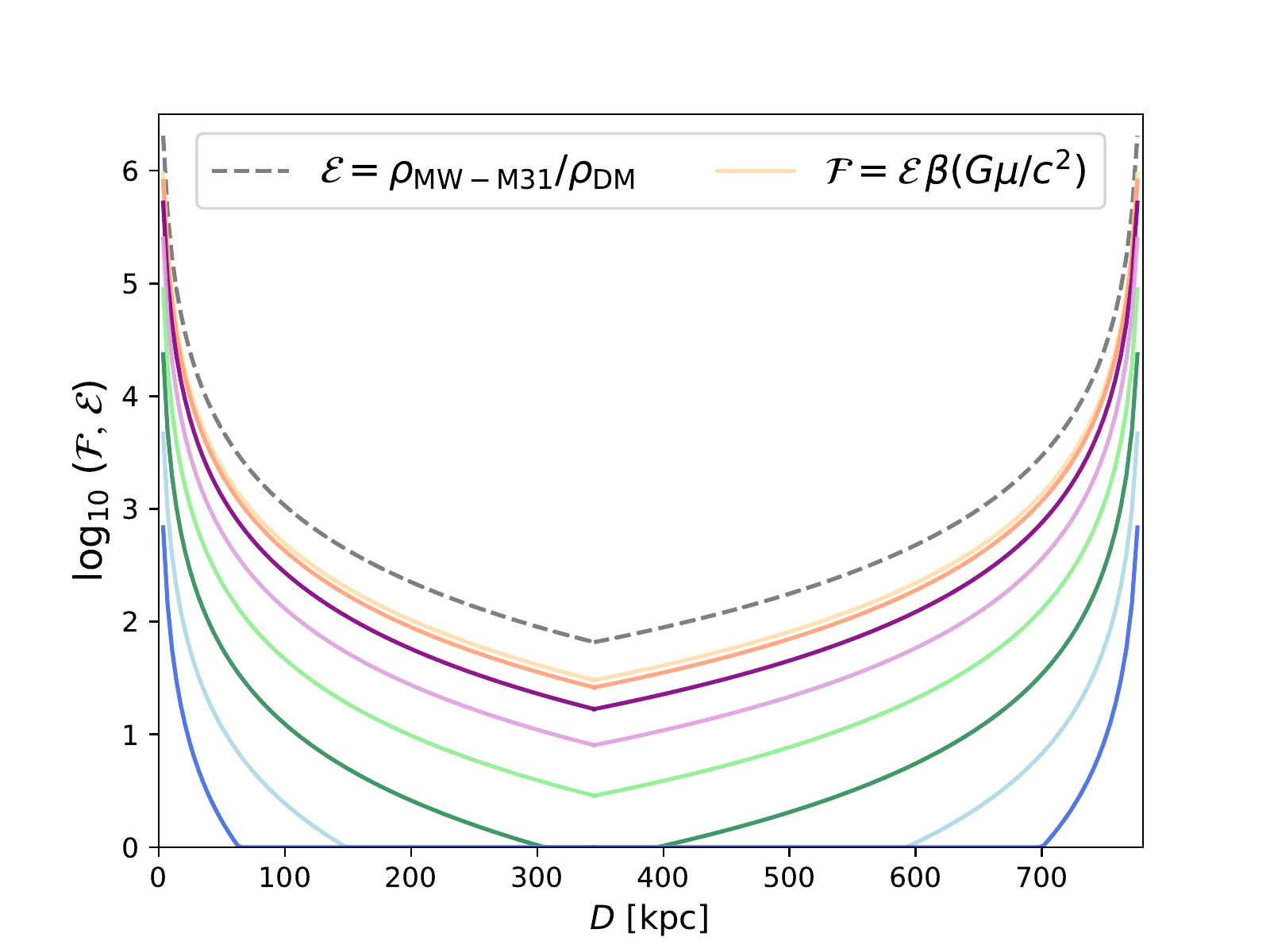}
     \hspace{0.05cm}
    \end{minipage}
 \caption{{\it Left panel:} The number density of string loops per logarithmic interval of length ${\rm d}\, n / {\rm d} \,\ln l$ as a function of loop size for string tensions $G \mu/c^2=10^{-8}$ to $10^{-15}$ (bottom to top) in powers of $10^{-1}$ and ${\cal G}=1$. This is the baseline distribution for a $\Lambda$CDM cosmology with ${\cal G}=1$ and no clustering ${\cal F}=1$. \Rp: DM enhancement (${\cal E}$, dashed line) and string enhancement (${\cal F} = {\cal E} \, \beta (\tension) $, solid lines) along the line of sight from the MW to M31 for varying string tensions (colours for the different values of $\tension$ as in the \lp). The maximum enhancement is about $10^5$ near the local position; the minimum value along the line of sight is ${\cal F} = 1$, the homogeneous limit. \label{fig:StringDMDistrib} }
\end{figure*}
%

The intrinsic homogeneous distribution of string loops $\dndloglinline$ for a $\Lambda$CDM cosmology today is shown in the \lp\ of \autoref{fig:StringDMDistrib} for a range of string tensions with no string theory enhancements and no clustering enhancements (CT18). We call this the baseline distribution $\dndloglinline_{\rm \, base}$. For each tension the peak in the distribution occurs near $l_g$. These distributions have many loops much smaller than typical galactic scales for $G \mu/c^2 \, \lta \, 10^{-9}$. An analytic approximation (CT18) for the number density per logarithmic interval for strings of loop length $l$ in the Universe today is
\begin{equation} \label{eq:dn_dlnl}
\left( \frac{\dd \, n}{\dd \, {\rm ln}\, l} \right)_{\rm base} = 1.15 \times 10^{-6} \frac{x}{(1+x)^{5/2}} \frac{f_{\, 0.2} \, \alpha_{0.1}^{1/2}}{\left(\Gamma_{50} \mu_{-13}\right)^{3/2}}  \,  \, {\rm kpc}^{-3} \, ,
\end{equation}
with $\mu_{-13} = \tension/10^{-13}$ , $x=l/l_g$ and gravitational length $l_g = 0.0206 \, \Gamma_{50} \, \mu_{-13} \, {\rm pc}$. These loops are cut from horizon crossing strings in the scaling regime, not closed strings created directly by the brane collision. The result has been written in terms of parameters $\Gamma = 50 \, \Gamma_{50}$ (the dimensionless rate for gravitational emission of loops); $f = 0.2 \, f_{\, 0.2}$ (the fraction of the network that forms large loops which are the loops of interest here); $\alpha = 0.1 \, \alpha_{0.1}$ (the size of the large loops relative to the horizon). The simulation determined network values are $\Gamma_{50}=\alpha_{0.1}=f_{\, 0.2}=1$. There are numerical uncertainties in $\Gamma$, $f$ and $\alpha$ but these are small compared to the intrinsic uncertainty in ${\cal G}$ and in string tension. We will refer to the values of $\Gamma$, $f$ and $\alpha$ as `simulation determined network values' and will not change them. This general treatment ignores non-gravitational channels for loop decay which might include axions or electromagnetic emissions. 

The relevant astrophysical distribution depends upon ${\cal G}$, string theory enhancements over SSSUIP, and ${\cal F}$, clustering enhancements over the mean in the universe. The homogeneous distribution (e.g. in the IGM) is:
\begin{equation} \label{eq:dn_dlnl_hom}
    \dndlogl_{\rm hom}={\cal G}\dndlogl_{\rm base}
\end{equation}
and the distribution in a collapsed structure is:
\begin{equation}  \label{eq:dn_dlnl_inhom}
\dndlogl_{\rm inhom}={\cal F} \dndlogl_{\rm hom}={\cal F} {\cal G} \dndlogl_{\rm base} .
\end{equation}
The homogeneous distribution is sensitive to ${\cal G}$ while the collapsed distribution is sensitive to the product ${\cal F}{\cal G}$.

The homogeneous string loop density in the Universe depends upon string tension. Integrating Eqs.~\ref{eq:dn_dlnl} and \ref{eq:dn_dlnl_hom} over all loop lengths gives the total density in the superstring loop distribution
\begin{equation}\label{eq:omega_homog}
\Omega_{\rm loop} = 2.4 \times 10^{-10} 
\left( \frac{ {\cal G} \, f_{0.2} }{h^2} \right)
\left( \frac{\alpha_{0.1} \, \mu_{-13}}{\Gamma_{50}} \right)^{1/2} 
\end{equation}
where the Hubble constant is $H_0 = 100 h$ km/s/Mpc. Experimental searches are challenging because $\Omega_{\rm loop}$ is small.

The tendency of superstring loops to cluster \citep{Chernoff:2009tp} may be summarised as follows: the preponderance of today's loops are formed before the epoch of radiation-matter equality for $G \mu/c^2 < 7 \times 10^{-9}\, \alpha_{0.1}/ \, \Gamma_{50}$. Slow moving loops cluster like cold dark matter when gravitational instability causes structure to form. The loops trace the same spatial distribution but contribute virtually nothing to the total mass density. In the case of the formation of our own Galaxy, loops of low tension start to cluster when $G \mu/c^2 \, \lta \, 4 \times 10^{-9}$ and their density enhancement ${\cal F}$ (ratio inside the Galaxy to the IGM) rises as $\mu$ parametrically decreases. Specifically, ${\cal F}$ increases as $\mu$ decreases in the range $10^{-13} < G \mu/c^2 < 10^{-9}$; ${\cal F}$ saturates for $G \mu/c^2 < 10^{-13}$. The enhancement at the solar position, ${\cal F} \sim 10^5$, approaches that of cold dark matter when $G \mu/c^2 \, \lta \, 10^{-13}$, these features can be seen in the \rp\ of \autoref{fig:StringDMDistrib}.  By contrast, in the mean IGM ${\cal F}=1$ for all string tensions $\tension$.

We model the inhomogeneous string loop distribution in a structure today by scaling its empirically observed dark matter density $\rhodm$. The spatially dependent enhancement of the local dark matter density is ${\cal E}=\rhodm/(\Omdm \, \rhocrit)$, where the denominator is the mean dark matter density in the universe, $\Omdm \, \rhocrit$, for critical density $\rhocrit$. The knowledge of the dark matter profile is sufficient to estimate the string loop profile: the string loop enhancement is ${\cal F} = \beta \, (\mu) \, {\cal E}$ where $0<\beta(\mu)<0.4$ has been fitted to numerical simulation (CT18). For $\beta=0.4$ the string loops cluster almost as strongly as cold dark matter; if $\beta$ is small the string loops are unclustered.

Roughly speaking, the detection rates of all experiments searching for loops scale with ${\cal F} {\cal G} \rsource$ where $\rsource$ is the typical length scale over which lensing occurs. There are two interesting circumstances where the product is large: (1) probing collapsed, galactic-scale structures with large ${\cal F}$ when $G \mu/c^2 < 10^{-9}$ and (2) monitoring very distant point sources (e.g., supernovae) with large $\rsource$ for any tension. We forecast the expected number of detections for superstrings in  both situations. The simulation derived network parameters are fixed and we explore the dependence on the string theory enhancement, $\cal G$, and the string tension, $\tension$. The modelling of the underlying DM and string loop distributions for the experiments considered in this work are explained next.


%
\subsection{Modelling of Dark Matter and String Loop Distribution} \label{sec:DensDistr}
%

The distribution of cosmic superstrings along the line of sight and its proxy, the DM density, have to be modelled to calculate event rates. Our assumptions for the two different observation targets are given below. 

\begin{itemize}
\item {\bf DM for MW-M31}: We assume isolated, spherical DM halo profiles for both MW and M31 as described by simple theoretical models of structure growth for self-similar, radial infall (for a review, see \citealt{Fich:1991ej}). Each individual profile applies outside the central region where baryons dominate and inside the tidal radius to a massive neighbor. For MW and M31 each profile plausibly applies to halocentric radii $\sim 5-300$ kpc. The joint MW-M31 model is fixed when we require (1) the density and mass of the profiles scale like $r^{-9/4}$ and $r^{3/4}$, respectively, (2) the density of the two profiles match at a point between the two centres such that the total (interior, spherical) mass of M31 is twice that of the MW, and (3) the MW rotation velocity is $220$ km/s at radius $r=8.5$ kpc. The DM density profile at distance $r$ from the MW centre is
\begin{equation}
    \rho_{\rm MW-M31} = \left\{ \begin{array}{c} \frac{A}{r^{9/4}} \,\, \, 0<r<r_1\\ \frac{2^{3/4}A}{(B-r)^{9/4}} \, \, r_1 < r < B \end{array} \right.
\end{equation}
where $B = r_1 + r_2$ is the distance to M31. For $r$ in kpc, $A = 1.15 \times 10^9 \, M_\odot$/kpc$^3$, $r_1 = 345$ kpc, $r_2 = 435$ kpc, $B = 780$ kpc \citep{1998ApJ...503L.131S} then the total masses are $1.54 \times 10^{12} M_\odot$ for the MW and $3.08 \times 10^{12} M_\odot$ for M31. In applications for stars directly behind the centre of M31 we avoid the diverging central density by introducing a $100$ pc minimum impact parameter for the line of sight. The efficiency of detection will not be greatly impacted by the choice. However, the detection rates are sensitive to the specific cutoff and we describe the effect of adopting a $1$ kpc minimum when we present results.

\item {\bf DM for IGM}: We assume a constant DM density $\rhodm = \Omdm \, \rhocrit$, with $\Omdm=0.25$,  $\rhocrit=3 H_0^2/(8 \pi G)$ and $h=0.7$. The enhanced DM densities in the MW and the DM halo of the host galaxy of the lensed sources are neglected in the simulations. Adding the host haloes into the calculation would lead to a higher string density close to the observer and source, giving rise to more microlensing events with shorter durations, see \autoref{sec:time-scales}. Neglecting these host halo events is purposeful as the intent is to establish the contribution of the cosmological, IGM-related line of sight. 

\item {\bf String loop distribution}: With a given DM density distribution we model the spatial string loop distribution. Low tension strings trace the dark matter more closely than high tension strings. This can be seen in the \rp\ of \autoref{fig:StringDMDistrib} which shows the string (${\cal F}$) and DM enhancement (${\cal E}$) along the line of sight for the M31 experiment.

\end{itemize}

%
\begin{figure*}
\begin{minipage}{0.45\linewidth}
     \centering
     \includegraphics[trim= 150.0 150.0 150.0 150.0, clip, width=\hsize]{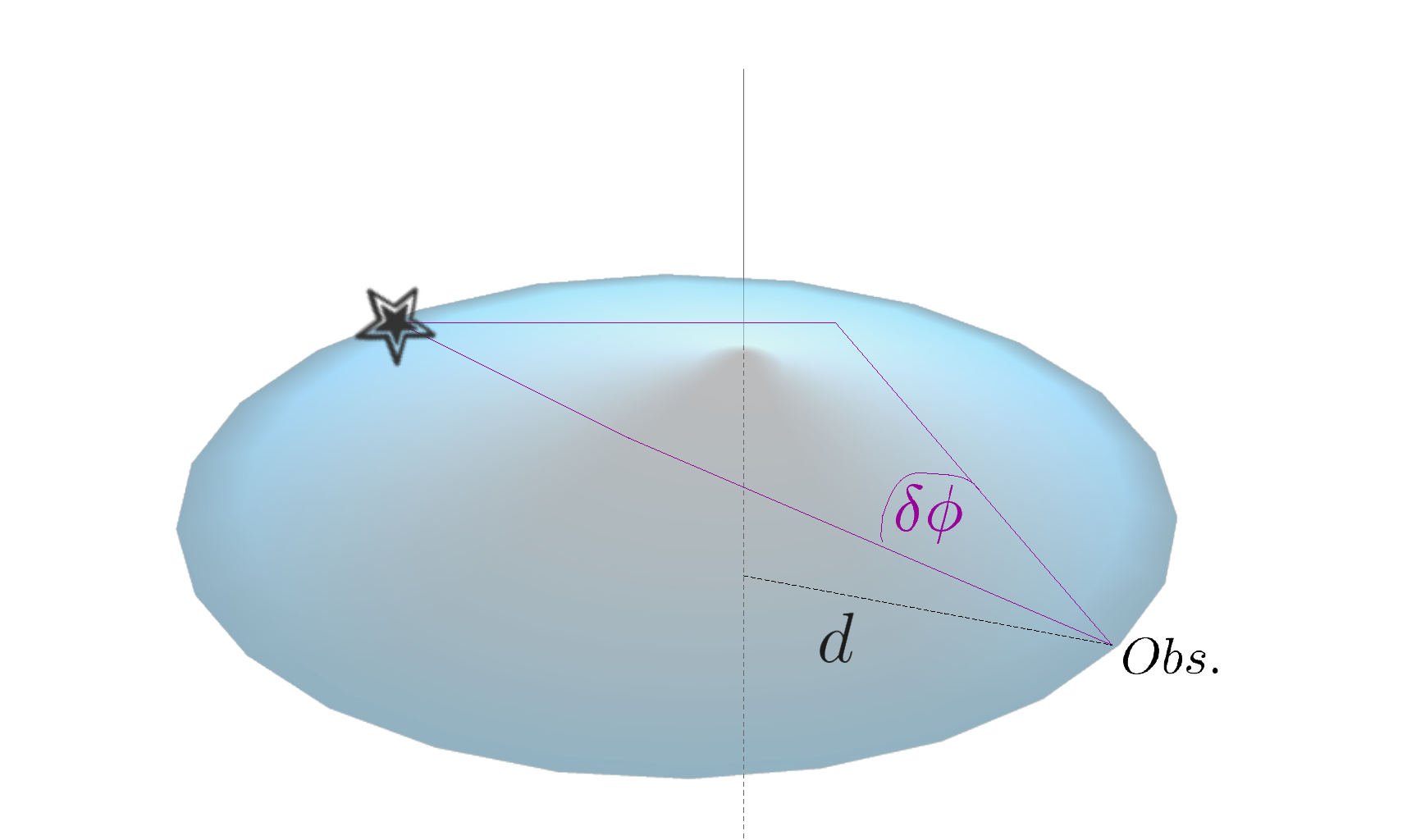}
     \vspace{0.05cm}
    \end{minipage}
    \begin{minipage}{0.45\linewidth}
      \centering\includegraphics[trim= 0.0 0.0 0 0.0 0, clip, width=7.9cm]{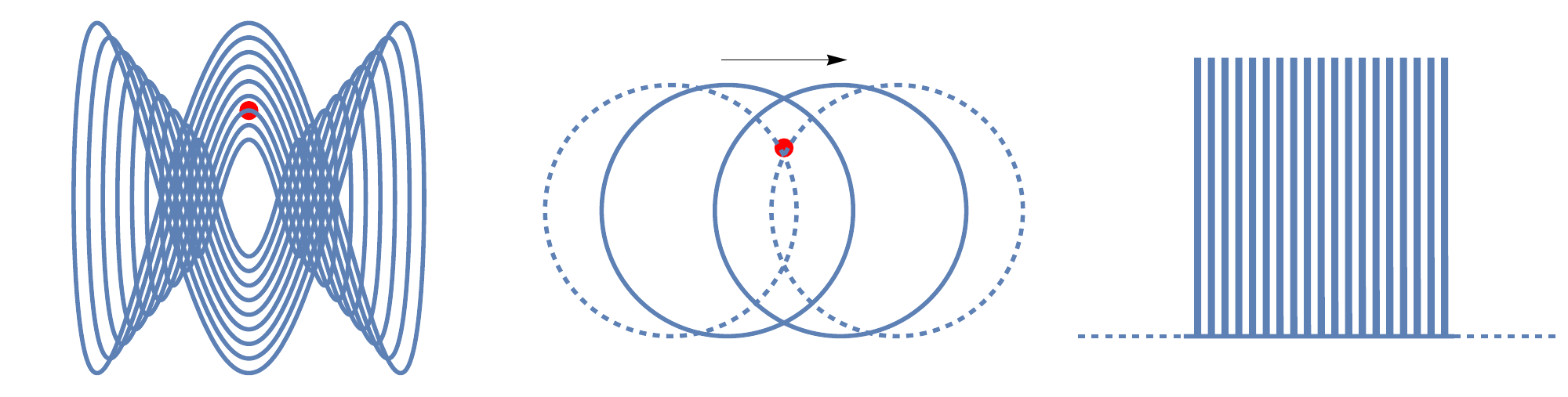}
     \hspace{0.05cm}
    \end{minipage}
 \caption{ \Lp: Geometry of a microlensing event; the string carves out a wedge of the disk perpendicular to the string with angle $\Delta \Theta = 8\pi G \mu/c^2$. The edges are identified and the resultant space is flat and conical. There are two straight line paths from a source to an observer located in a specific region directly behind the string. When the string (at rest with respect to the observer) lies perpendicular to the line of sight the observer sees two images of the star separated by angle $\delta \phi = \Delta \Theta \,   \big( 1-\frac{\rstring}{\rsource}\big)$, where the distance from the observer to the string is $\rstring$, and the distance from observer to source is $\rsource$. When the string lies at angle $\thetastr$ with respect to the line of sight the splitting is diminished according to $\Delta \Theta \to \sin \thetastr \Delta \Theta$. 
 \Rp: Schematic renditions of moving loop (\emph{blue}) and stationary source (\emph{red}). The left picture shows snapshots of the internal configuration of the loop oscillating with time-scale $\tosc$. The middle picture shows the projected area covered by the oscillating loop as its centre of mass proceeds in the direction of the arrow. The time-scale for the interaction of source and loop is $\tpass$. The dotted lines show the beginning and end of the time of passage or the `geometric alignment'. The right picture shows the train of microlensing events that results. The flux is 1 (\emph{lower}) and 2 (\emph{upper}) times the nominal source value. 
 \label{fig:geom} }
\end{figure*}
%

%
\section{String Microlensing} \label{sec:ML}
%

Unlike a Newtonian point mass which curves space, a straight string's positive energy density and negative pressure (along its length) conspire to leave space flat. Consider the geometry shown in \autoref{fig:geom}, with a string perpendicular to the line of sight. The string induces a deficit angle $\Delta \Theta= 8 \pi G \mu/c^2$ and a conical geometry.  There is no magnification, shear or distortion of a particular image, but there are multiple distinct paths for photons to travel from source to observer when the source, string and observer are all nearly aligned. Let the distance from the observer to the source be $\rsource$
and from the observer to the string be $\rstring$.

A straight string at rest in the observer's frame generates an angular
separation of lensed image pairs $\left( \delta \phi \right)_{u=0} = \Delta \Theta
|\sin \thetastr| (1-\rstring/\rsource)$ where $\thetastr$ is the angle
between the string tangent vector and the line of sight
\citep{Vilenkin:2000jqa}. When the string moves with respect to the
observer the splitting is $\delta \phi = \left( \delta \phi \right)_{u=0} 
  \left(\gamma^2 \left( 1 + u_n/c \right)^2 - \cos^2 \thetastr\right)^{1/2}/|\sin
    \thetastr|$ where $u$ is the string velocity (perpendicular to the string; $\gamma=1/\sqrt{1-u^2/c^2}$) and
$u_n$ is the component along the line of sight to the source
  \citep{Shlaer:2005gk,Shlaer:2005ry}. For small velocities $\delta
  \phi/\left(\delta \phi\right)_{u=0} = 1 + u_n/(c \sin^2 \theta)$. In this work we
  adopt the static limit ($u/c=0$) for the size of the angular
  splitting.

The scale of the angular splitting of images for GUT strings with $G\mu/c^2 = 10^{-6}$ is $\Delta \Theta \sim 5$ arc-seconds. Exact double images of cosmic objects like galaxies located behind the string can potentially reveal the string's presence. Since that suggestion was first made, the observational bounds (from CMB, gravity wave searches and pulsar timing) have tightened, certainly constraining $G\mu/c^2 \lesssim 3 \times 10^{-8} - 3 \times 10^{-7}$, so that the expected splitting cannot be resolved. In addition, string theory understanding of the problem of compactification naturally yields superstrings, string-like entities with much smaller string tension. In this context, one looks for string microlensing \citep{Chernoff:2007pd,Bloomfield:2013jka}, the transient change of the unresolved flux when a star passes behind the string. Even though the details of the light bending differ, such a search is conceptually similar to microlensing searches for dark Newtonian masses.

String microlensing is sudden whereas Newtonian microlensing is continuous.\footnote{Colloquially, string microlensing is {\it digital} whereas Newtonian microlensing is {\it analog}.} The brightness of point-like lensed sources will appear to fluctuate achromatically {\it by a factor of 2} as the angular region associated with the string passes across the observer-source line of sight. The point-like limit is satisfied when the angular scale of the source is small compared to the deficit angle, as will be described quantitatively in \autoref{sec:ObsLimits}. The segments of the string oscillate with relativistic velocities. Hence, the characteristic duration of the event is 
\begin{equation} \label{eq:time-scale}
\tlens \sim \frac{\rstring \,  \Delta \Theta}{c} \sim 630 {\rm \ s \ } \frac{\rstring}{10 {\rm \ kpc \ }} \, \frac{G \mu/c^2}{2 \times 10^{-10}} \, ,
\end{equation}
where $\rstring$ is the characteristic distance of observer to string. The total rate of lensing of a source ($\Rlens$) by a distribution of loops $\dndlinline$ is proportional to the solid angle swept out per time $\sim c l/r^2$ for a loop $l$ at distance $r$  or 
\begin{equation} \label{eq:Rlens}
    \Rlens \sim \int  \dd \, l \, \dd \, r \, r^2 \dndl \frac{c l}{r^2} \, .
\end{equation}
The distribution $\dndlinline \propto l^{-2.5}$ for the small loops made prior to matter-radiation equality that have not yet evaporated because $l \, \gta \, l_g$.  The first moment of the distribution is dominated by the gravitational cutoff $l \sim l_g \propto \mu$. Consequently, $\Rlens \propto 1/\sqrt{G \mu/c^2}$ and smaller tensions give larger lensing rates.

By contrast, the probability of lensing at any instant of time is
\begin{equation} \label{eq:Plens}
    \Plens \sim \int \dd \, l \, \dd \, r \, r^2 \dndl \frac{l \Delta \Theta}{r}
\end{equation}
which yields $\Plens \propto \sqrt{G \mu/c^2}$. Experiments are sensitive to the static probability of lensing at large tension and the rate of lensing at small tension.

String microlensing has several additional distinctive features. The internal motions of a string loop are relativistic but when a loop is bound to a structure like the Galaxy then its centre-of-mass motion is $\vcom \sim 220$ km s$^{-1}$. Microlensing of a given source will {\it repeat} $\Nrep \sim c/\vcomperp$ times (where $\vcomperp$ is the centre-of-mass velocity perpendicular to the line of sight). In such a case the characteristic loop oscillation time-scale governs the repetition time-scale for a single source, $\tosc \sim l_g/c = 0.067 {\rm \ yr \ } \mu_{-13}$ whereas the centre of mass velocity governs the time-scale for the passage, $\tpass \sim l_g/\vcomperp \sim 90 {\rm \ yr \ } \mu_{-13} $. The microlensing that occurs during the time of passage (hereafter `alignment') is schematically shown in the \rp\ of \autoref{fig:geom}.

For strings of invariant length $l$ the fundamental period of oscillation in the string centre of mass is $l/(2 c)$. The string's internal rms velocity (averaged over the length and the fundamental period) is $c/\sqrt{2}$. The magnitude of the internal string velocity perpendicular to the line of sight (perpendicular to the loop's tangent vector, averaged over loop orientation, length and period) is $|v_{\perp}| \sim 0.5 c$, a result that depends weakly on the details of the loop's shape. For estimating time-scales of microlensing we fix $|v_{\perp}|=0.5 c$ and, where necessary, we also assume that the angle of the string with respect to the line of sight is $\thetastr=\pi/4$. Selecting specific values is an approximation which neglects the spread owing to the range of orientation geometries and the detailed oscillation dynamics of the loop. 

The relative motion of the loop centre-of-mass and of the line of sight to the source in M31 on the sky is $\vcomperp$.  We account for M31's heliocentric motion and direction, the random motion of the source with respect to the centre of M31 (assumed isotropic with 1 dimensional dispersion $\sim 150$ km/s) and the random lens motion in the halo (assumed isotropic with 1 dimensional dispersion $\sim 100$ km/s at midpoint between the Galaxy and M31). Depending upon exactly the position of the lens's crossing along the line of sight we estimate $\sqrt{<\vcomperp^2>}$ varies from  $140-300$ km/s. The minimum is roughly half-way between the two galaxies. Anticipating that the longest microlensing events are of greatest observational interest we select $150$ km/s for the typical size of $\vcomperp$ for the M31 study.

The lower bound on the relative motion of loop and line of sight to a supernova ignores the rocket effect. It is estimated from three inputs: the Sun's heliocentric velocity with respect to the CMB ($\sim 369.5$ km/s, \citealt{Kogut:1993ag}), the lens's motion in the IGM and the peculiar velocity of the SN~Ia. The velocity of the lens is taken to be identical to that of the dark matter in $\Lambda$CDM cosmology (assumed isotropic with 1 dimensional dispersion $\sim 280$ km/s, \citealt{Bahcall:1994mj}). The peculiar velocity of SNe~Ia may be estimated from the residual errors in the calibration of light curves in low redshift samples ($\sim 150 - 250$ km/s after all systematic effects are accounted for, \citealt{Stanishev:2015eva}). Alternatively, it may be estimated from the peculiar velocity of peaks in $\Lambda$CDM cosmology (a linear analysis of gaussian random fields $\sim 290$ km/s, \citealt{Suhhonenko:2002wb}).  Varying the distance to lens, direction of the source with respect to the Sun's motion in the frame of the CMB and the assumption of peculiar SNe~Ia we find that $\sqrt{<\vcomperp^2>}$ ranges over $400-700$ km/s. We take $600$ km/s to be the lower bound for $\vcomperp$. This ignores the fraction of SNe~Ia that occur in rich clusters where the cluster dispersion can be comparable to or larger than this value.

The upper bound assumes the maximal rocket effect. A loop of
characteristic size $\ell_g$ today can reach speeds $\sim 0.04 \, c$ in a
$\Lambda$CDM cosmology assuming it was created by intercommutation
with speed $\vinter/c=0.1$, size $\alpha=0.1$, spent its entire life
in the IGM and was accelerated in a fixed direction with maximum
asymmetry (10\%).\footnote{The speed depends modestly upon the choice
  of loop size: the peak of the loop-weighted density occurs at
  $(2/3)\, \ell_g$ and gives velocity $\sim 0.06 \, c$.} We estimate the
upper bound $\sqrt{<\vcomperp^2>} \sim 9,800$ km/s which is $\sim 16$ larger
than the lower bound.

To summarize, $\sqrt{<\vcomperp^2>} \sim 150$ km/s for the M31
study while spanning the range $600 - 10,000$ km/s for the IGM study.  The
translational speed of the loop center of mass is generally smaller than
the internal oscillatory motions (rms velocity
$c/\sqrt{2}$).  The internal velocity sets the duration of a single
microlensing event, the center of mass velocity sets the total time of
alignment (i.e. the duration of a series of microlensing events) and
the number of events in a series is the ratio of two velocities.

%
\section{Observational Limits} \label{sec:ObsLimits}
%
In general, there are several important limitations on observing an exact factor of two enhancement in flux during lensing: (1) the source must be point-like in the sense that its angular size is small compared to $\Delta \Theta$, (2) the measurement interval for the flux determination must be less than the duration of the microlensing event, (3) the signal to noise must be sufficiently large that false detections and false dismissals are negligible and (4) the telescope's angular resolution must be high enough to avoid contaminating the flux of the point-like object with unresolved background light.

Previous studies of microlensing of stars by strings focused on Galactic stars \citep{Chernoff:2014cba}. Here we explore the feasibility to detect superstring microlensing with extragalactic sources. In particular, we will consider the lensing of stars in Andromeda and of supernovae at cosmological distances.

Andromeda (M31) is our nearest galaxy neighbour beyond the Magellanic Clouds, yet, the entire galaxy can be fitted within the field of view of the instruments in ongoing time-domain surveys, e.g., the Zwicky Transient Facility \citep{2019arXiv190201945G}. Single stars with radius $\sim R_\odot$ in uncrowded fields in M31 are point-like for $G \mu/c^2 \, \gta \, R_\odot/(8 \pi \rsource) \sim 10^{-15}$.  The rates benefit from the large product ${\cal F}{\cal G}$ (see \autoref{sec:cosmo} for details). We have explored the effect of cadence and exposure time of the observations on the expected number of microlensing detections per star.

For the microlensing of supernovae, we assume that the superstring loops are homogeneously distributed throughout the IGM. Supernovae have the advantage that they can be observed at much further distances than stars and contamination effects are minimal. With typical supernova shell sizes $R_{\rm sh} \sim 10^{-2}$ pc, and $\rsource \sim 1$ Gpc distances, supernovae can be treated as point sources for string tensions $G \mu/c^2 \, \gta \, R_{\rm sh}/(8 \pi \rsource) \sim 4 \times 10^{-13}$. The rates benefit from large $\rsource {\cal G}$. The microlensing of stars in M31 and supernovae at cosmological distances are potentially complementary in probing the parameter space of strings. 

Finally, we note that experiments with quasars (QSOs) and active galactic nuclei (AGNs) have two major difficulties. First, QSOs with size $R_{\rm QSO} \, \gta \, 1$ light year (based on time variation of optical emission) at Gpc distances are effectively point-like for $G \mu/c^2 \,\gta\, 10^{-11}$. Cosmological microlensing experiments utilising lenses in the IGM, described by $\dndlinline_{\rm \, homog}$, achieve reasonable rates of microlensing only for tensions smaller than this point-like cutoff.\footnote{\cite{2008MNRAS.384..161K} estimated the source size cutoff in $G \mu/c^2$ to be a few times larger than the number quoted above. They modeled a lensing distribution similar to $\dndlinline_{\rm \, base}$ and concluded that rates for strings with $G \mu/c^2 \, \gta \, 4 \times 10^{-11}$ were too small to be detectable. Even though our lensing distribution is larger by the factor ${\cal G}$ the same difficulty stands in the way of making use of such sources.} Second, the intrinsic source fluctuations of QSOs and AGNs make it problematic to identify a factor of two change in flux.

%
\section{time-scale of Events} \label{sec:time-scales}
%

The time-scale of microlensing is a key consideration for experiments hoping to detect superstring microlensing events. Here we compare the expected time-scales for observations of M31 and distant supernovae.

%
\begin{figure*}
\begin{minipage}{0.45\linewidth}
     \centering
      \centering\includegraphics[trim= 0.0 0.0 20 20.0 0, clip, width=7.9cm]{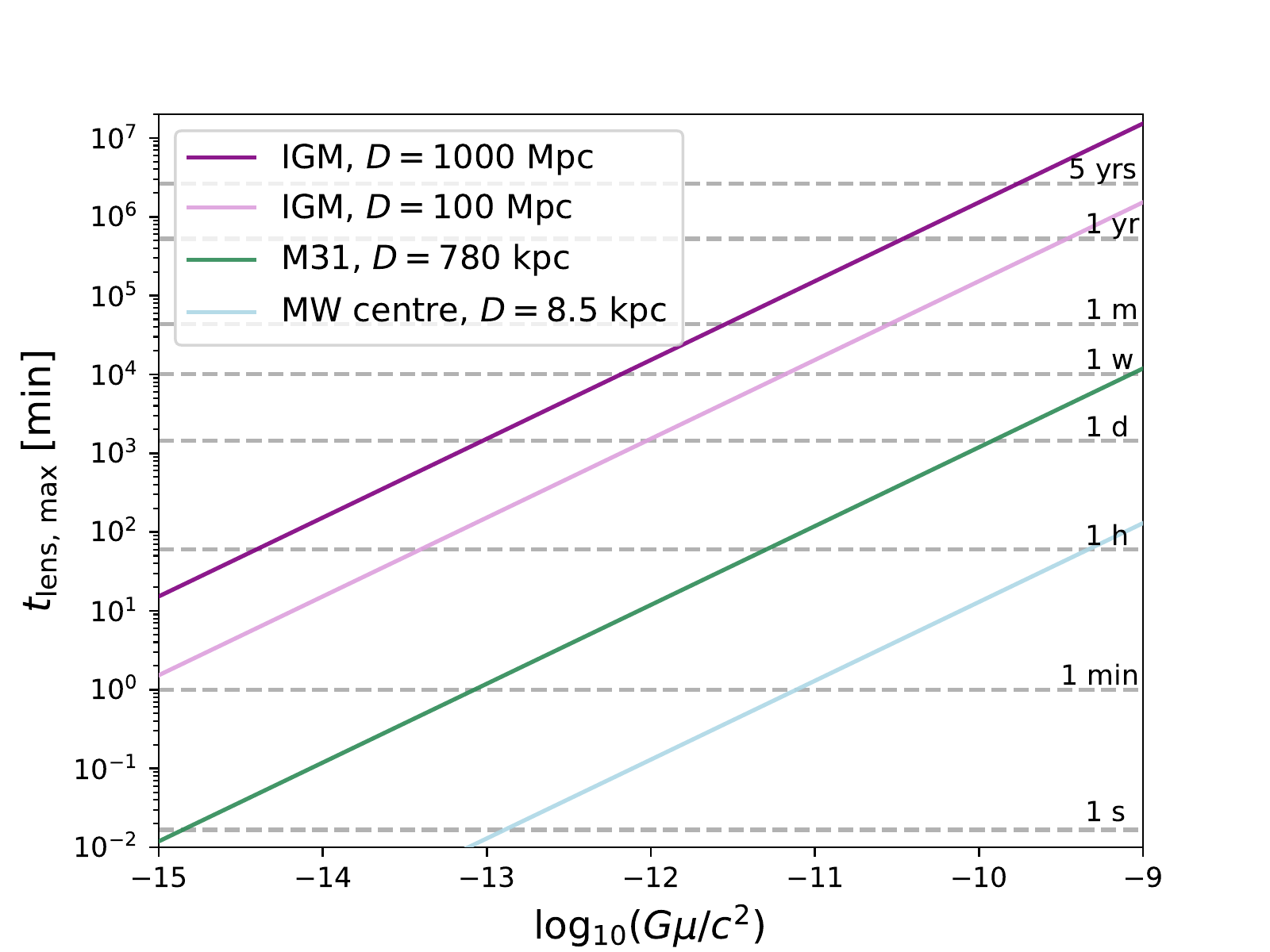}
     \vspace{0.05cm}
    \end{minipage}
    \begin{minipage}{0.45\linewidth}
     \includegraphics[trim= 0 0 20 20 0, clip, width=7.9cm]{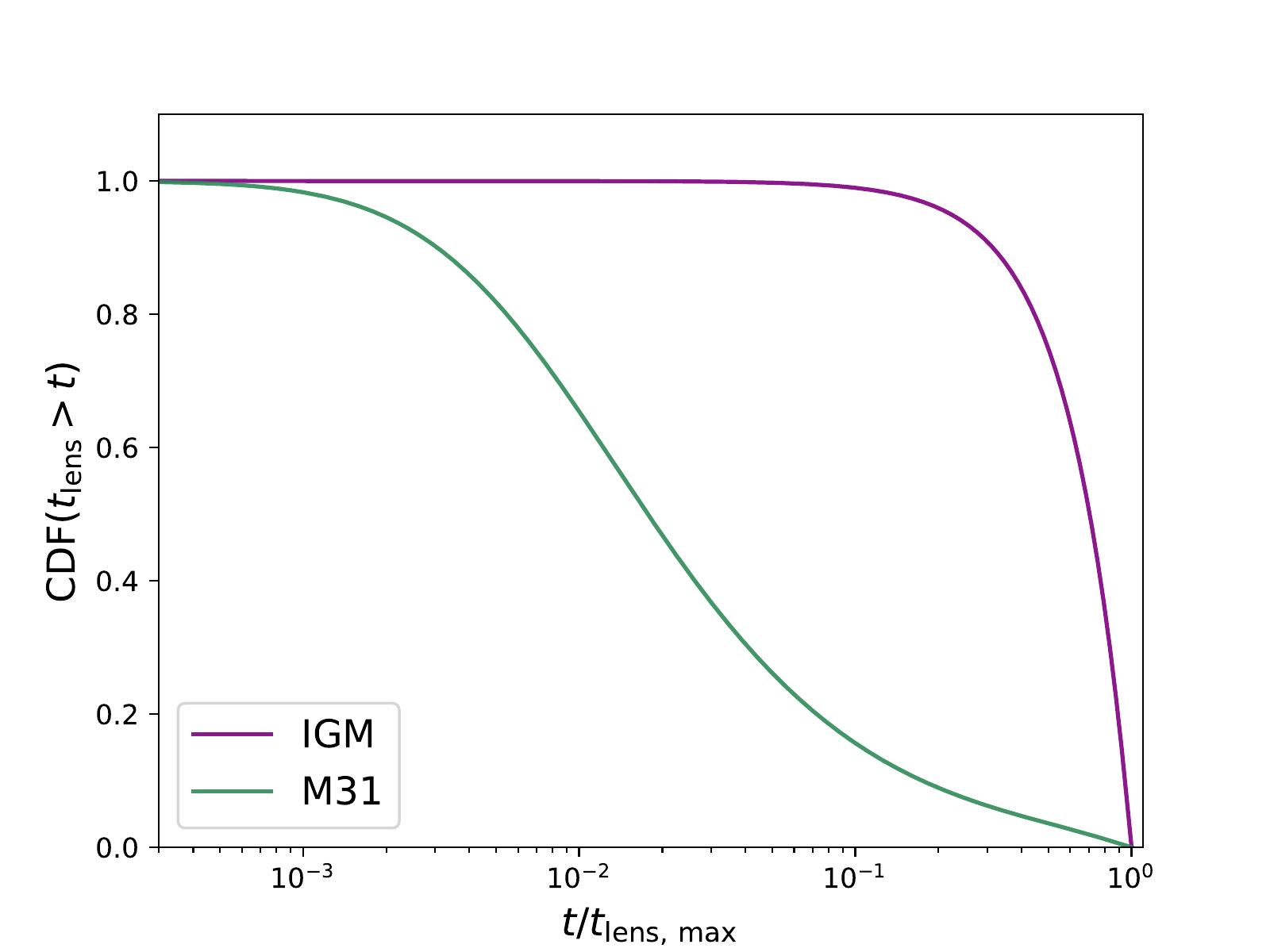}
     \hspace{0.05cm}
    \end{minipage}
 \caption{{\it Left panel:} The maximum lensing time depending on the string tensions and distance to the lensed source. \Rp: The scaled cumulative probability distribution function (CDF) for the duration of a microlensing event of a star in M31 $8$ kpc in front or in the plane of the galaxy (\emph{green line}) and of a distant source lensed by a cosmic superstring in the IGM (\emph{purple line}) (the scaled IGM graph is distance independent). The abscissa is $\tlens/\tlensm$ where the maximum lensing duration is $\tlensm$, given by Eq.~\ref{eq:time-scales}. \label{fig:time-scales}}
\end{figure*}
%

\subsection{Microlensing of Stars in near-by Galaxies: the Case for Andromeda} \label{sec:ML_M31}

To deduce the time-scale sensitivity we simulate an experiment at a prescribed set of epochs for observations of a source located at a fixed position in M31. We assume that lensing of individual stars can be treated as statistically independent. The lensing strings are drawn from a model based on the density distribution $\dndlinline_{\rm \, inhomog}$ for an assumed string tension (with ${\cal G}=10^2$). The procedure depends upon the line of sight that passes from the observer, through the MW and M31 haloes, to the source. We calculate the overlap of exposure windows and microlensing epochs and infer the average flux for each exposure. Thereafter, we apply plausible detection criteria to the sequence of fluxes measured during the course of the full experiment. As discussed in \autoref{sec:simTech} the expected number of detections per source for the full experiment depends upon various survey characteristics including the number of exposures, the time of each exposure, the total length of observations, etc. We estimate the number of stars needed to observe microlensing for a given string tension in a range of experiments with given survey characteristics.

An important quantity is the typical duration of the microlensing event for a given source. `Event' means a single transient doubling of the flux of the source (see the \lp\ of \autoref{fig:geom} for the configuration of the observer, source and a segment of the loop). An event may only occur during an `alignment', the period that the line of sight from observer to source passes through the loop's period-averaged projected area on the sky.\footnote{See the middle picture of the right panel of \autoref{fig:geom}. The period of geometric alignment is the time of passage, the time it takes the loop's time-averaged, projected area to clear the line of sight. It subsumes $\sim c/\vcomperp$ individual microlensing events and the intervening unlensed time intervals as shown in the right picture of the right panel of \autoref{fig:geom}.} There are typically $\Nrep$ events per geometric alignment as the relativistically moving segment of the oscillating loop passes repeatedly through the observer-source line of sight (see \rp\ of \autoref{fig:geom}).
The duration of one microlensing event is
\begin{equation} \label{eq:time-scales}
\tlens = \frac{8 \pi G \mu}{ c^2} \, \sin \thetastr\frac{\rstring}{v_\perp} \bigg( 1 - \frac{\rstring}{\rsource} \bigg) \, ,
\end{equation}
where $\rstring$ is the distance between observer and string, $\rsource$ is the distance from observer to source, $\thetastr$ is the angle of string to the line of sight and $v_\perp$ is the velocity of the string perpendicular to the line of sight in the instantaneous rest frame defined by the line of sight. Typically, the string's relativistic oscillatory motion dominates $v_\perp$. 

The lensing duration varies linearly with $\rstring$ at fixed $\rstring/\rsource$. The event time-scale shrinks when the string lens is close to the source or close to the observer and the longest events at fixed $\rsource$ have $\rstring = 0.5 \, \rsource$, 
\begin{equation} \label{eq:time-scalesm}
\tlensm = \frac{8 \pi G \mu}{ c^2} \, \sin \thetastr\frac{\rsource}{4 v_\perp} \, .
\end{equation}
The resulting maximum time-scales as a function of string tension for different distances between source and observer are shown in the \lp\ of \autoref{fig:time-scales}. The scaling of this relation implies that M31 stars have characteristically longer microlensing event durations than Galactic stars at fixed string tension. Conversely, at fixed duration, M31 experiments are sensitive to lower string tensions than Galactic experiments.

The assumed underlying dark matter distribution (see \autoref{fig:StringDMDistrib} and \autoref{sec:DensDistr}) is scaled to give the corresponding loop distribution. We sample fairly the rate at which strings along the line of sight lens the background source to determine the distribution of event durations. We construct the cumulative probability distribution function (CDF) of durations for each string tension and source position. For the application to M31 it turns out that the CDF scales approximately with the maximum lensing time-scale for different tensions at fixed source position. In addition, a single scaled form of the CDF turns out to be accurate for source positions anywhere on the facing hemisphere, $8$ kpc from the centre of M31.
 
The \rp\ of \autoref{fig:time-scales} shows the scaled CDF for the M31 experiment. Cosmic superstrings cluster with DM, hence, most of the lensing events are generated by strings in dense central regions of the MW or M31. These are close to the source or close to the observer and give rise to short event durations ($\tlens < 0.01 \, \tlensm$). The typical lensing events will therefore last shorter than 1 s for small string tensions ($G\mu/c^2 \sim 10^{-15}$) making a detection of the events challenging. 

There are benefits to monitoring both M31 and MW bulge stars (see \lp\ of \autoref{fig:time-scales}). If we assume that CCD technology limits exposure times from a few seconds to an hour then exposures of M31 stars naturally probe the tension range $10^{-13.5}$ to $10^{-11}$ whereas exposure of MW bulge stars probe the range $10^{-11.5}$ to $10^{-9}$. These two experiments are complementary in sensitivity and sky coverage. Longer lines of sight to M31 make the requirements on exposure and cadence less demanding at small, fixed tension compared to the MW bulge. Observations of M31 can be carried out from the Northern hemisphere by large ongoing time-domain astronomy surveys while Southern hemisphere surveys can access the MW bulge.

\subsection{Microlensing of Supernovae by Cosmic Superstrings in the Intergalactic Medium} \label{sec:ML_SNe}

%
\begin{table*}
\caption{Parameters and assumptions (as defined in \autoref{sec:simTech}) for the simulation of the number of detected microlensing events in two surveys. The time $\tobs$ is the exposure time, $\Delta \tobs$ is the separation between the end and beginning of successive nightly exposures; $\delmi$ ($\delma$) quantifies the minimum upwards (maximum downwards) fluctuation of the measured flux used for the classification of a microlensing event detection (see main text for details). The first column is for stars in M31 lensed by cosmic superstrings in the haloes of MW or M31. The second column is for supernovae lensed by strings in the IGM.
}
\label{table:SurvSetUp}     
\centering                       
\begin{tabular}{c |c c }        
\hline\hline            
            & Stars in M31  &  Supernovae \\ \hline 
Source      & source at 8 kpc distance from M31 centre,                     & SNe at distances from $100$ Mpc  \\     
            & 6 positions for each distance, see \autoref{fig:analog_pos}   & up to 2000 Mpc \\ \\ 
Survey      
& Observed over 1 year for 2 hours per night
& Observed over 3 months for 5 times per night \\
& with repeated $1$s exposures separated by $0.01$s
& with repeated $30$s exposures separated by $1$ hour \\ \\
            
Min./Max. flux variation  & $\delmi = \delma = 30$ \%                           &  $\delmi = \delma = 10$ \%\\
for detection &           & \\ \\
\hline \hline                                   
\end{tabular}
\end{table*}
%

We also consider microlensing events of point-like, cosmological "standard candles" which are visible at Gpc distances. The time-scales of these events are much longer compared to the lensing of objects in the neighbouring environment of the MW since the maximum duration of a lensing event, $\tlensm$, scales with distance. The \lp\ of \autoref{fig:time-scales} shows $\tlensm$ at $100$ and $1000$ Mpc for a range of tensions. Note that long durations are anticipated for a considerable range of tensions below the current upper limits. 

The lensing is primarily sourced by the loops distributed in the intergalactic medium (IGM). The scaled CDF in the \rp\ of \autoref{fig:time-scales} shows that the distribution of event durations for the IGM has weight at larger $\tlens/\tlensm$ than in the case for M31. In the IGM there is no clustering and, to the extent cosmological evolution is neglected, the events are uniformly distributed along the line of sight and the scaled CDF is distance-independent. If $\tlensm$ is much bigger than the total duration of the experiment then catching the characteristic rise and fall of the flux due to the string crossing is improbable.

Type Ia supernovae (SNe~Ia) are particularly useful standard candles since they possess large luminosities, well-calibrated, homogeneous light curves and can be tracked for months after discovery \citep[see][for a review of the use of SNe~Ia as distance indicators in cosmology]{2011ARNPS..61..251G}. Any SN~Ia light curve that is a factor of two brighter than that expected for the source's redshift is easily identifiable. Standard candles, in general, and SNe~Ia, in particular, can be employed to detect and/or constrain cosmic superstring lensing at large tensions for which the characteristic duration of the event is long. Such observations have the capacity to discern even a constantly microlensed source.

This capability does not preclude detecting a microlensing event over a more limited time frame by seeing the flux double and then return to normal. While this would clearly be the preferred signature for a secure detection, \autoref{fig:time-scales} shows it is feasible only for smaller string tensions when $\tlensm$ is not too large. Of course, in that limit, all kinds of transients with smooth light curves, including all known supernova types, are useful. 

It is important to note that the point-source approximation can only be applied for string tensions $\tension \gtrsim 8 \times 10^{-13}$.\footnote{With slightly greater accuracy, consider a supernova at distance $\rsource$ with a spherical shell expanding at $10,000$ km/s at time $t$ after the explosion. For the angular diameter of the shell to be smaller than $\Delta \Theta$ we require $\tension \gtrsim 8 \times 10^{-13} (t/{\rm yr})({\rm Gpc}/\rsource)$. } For lower tensions the deficit angle becomes comparable to the size of the object and the flux change is less than a factor two.

We have neglected several cosmological effects in discussing the distribution of event durations. The IGM's string loop density changes with the look back time and this effects the relative rate of lensing by nearby and distant loops. A more accurate treatment would include not only the $(1+z)^3$ volume density change but also the intrinsic changes in the characteristic string loop size distribution since the peak of the length distribution is smaller at earlier times. We also neglected the redshift variation of time-scales (the event duration and the oscillation period) and of the angular diameter of the source. We do not expect any of these to make a qualitative difference over the modest redshift range of interest but we intend to return to consider them in more detail at a later date. 

A practical advantage of using SNe~Ia for the search for strings is that the standard observational strategy for cosmology need not be changed. The lensing time-scales are such that the data taken by ongoing and future surveys can be used to constrain the parameter space and possible events should be clearly detectable in the light curves.

%
\section{Event Rate Calculations and Efficiency Simulations} \label{sec:simTech}
%

The intrinsic physical microlensing event rate of stars with density $n_\star(r)$ by string loops with density length distribution $\dndlinline'$ (see \autoref{sec:cosmo}, Eq.~\ref{eq:dn_dlnl_inhom}) at position $r'$ is
\begin{equation} \label{eq:Gamma}
    \Gammalens = \int n_\star r^2 \, \dd \, \Omega \, \dd \,r \int \dd \, r' \, \dd \, l \,  \dndl'\left< \frac{\dd \, A_\perp}{\dd \, t} \right> \, ,
\end{equation}
where the observer is at the origin and lensing strings lie between observer and source ($0<r'<r$). The loop invariant length is $l$, the instantaneous rate at which the string loop covers area perpendicular to the line of sight $\left< \dd \, A_\perp/\dd \, t\right>$, averaged over the motion and orientation of the loop in spacetime and over the characteristic trajectories traced by different loops. In order of magnitude $\dd \, A_\perp/\dd \, t \sim l c$ since the loop oscillations are relativistic.  If $N_\star$ stars lie in a narrow cone at fixed distance $\rsource$ the microlensing rate per star per second is
\begin{equation} \label{eq:Gamma_perStar}
    \frac{\Gammalens}{N_\star} =  \int_0^\rsource \dd \, r' \dd \,  l \,  \dndl'\left< \frac{\dd \, A_\perp}{\dd \, t} \right>  .
\end{equation}
This quantity is proportional to the string theory enhancement parameter $\cal G$ and $\mu ^{-1/2}$ via $\dndlinline$ (Eq.~\ref{eq:dn_dlnl_inhom}). To use this formalism in the case of a single supernova set $N_\star \to 1$. The rate of alignment per source (passage of the time-averaged loop shape over the source) is simply $(\Gammalens/{N_\star})/\Nrep$.

An ideal experiment detects the intrinsic physical rate. To do so it makes precise, individual measurements with exposure times less than the time span of a single microlensing event, has no gaps in coverage and continues for a total duration much longer than an individual microlensing event. 
Actual experiments may fail to satisfy one or more of these conditions. Since the microlensing time-scale is a strong function of the string tension even a single experiment, near-ideal for a particular range of tensions, inevitably falls short over a wider range of tensions. In addition, many practical factors will mitigate against ideal design. We express the impact in terms of an efficiency $\epsilon$ such that $ \epsilon \, \Gammalens $ is the detectable rate.

We calculate $\epsilon$ for different experiments and for assumed values of the string tension by simulation. The string tension sets the deficit angle and, hence, the time-scale for the duration of one microlensing event at a given distance. The tension also influences the spatial density of string loops in collapsed structures and hence the probability of lensing. The exact experimental set-up controls the detection rate of events. The relationship between microlensing rate, alignment rate, event rate and detectable event rate is described in detail in \hyperref[app:efficiency]{Appendix~\ref*{app:efficiency}}. 

There are three main aspects to consider:

\begin{enumerate}[i)]

\item \textbf{Source position}: The position of the chosen source sets the DM density and the string distribution along the line of sight; when a lensing loop is selected the time-scale of the event is determined by the geometry of source, observer and loop. 

\item \textbf{Observation strategy}: The exposure time is crucial for determining whether a survey is sensitive to microlensing events of a given source. If the typical lensing time-scales are shorter than the exposure time, one can never resolve a single event and the flux amplification will be smaller than a factor of two. 

\item \textbf{Detection criteria}: What properties of a sequence of flux measurements are required for it to be called a detection? We have investigated several criteria for detection: \\
\underline{lens}: full pulse, i.e. a \emph{rising edge} (followed by consecutive enhanced flux measures) and a \emph{falling edge} are observed.%
\footnote{A rising edge is the observation of two consecutive flux measures where the first flux is low while the second is high. Accordingly, a falling edge is the reverse. The meaning of high and low depend upon thresholds.} %
Here, the digital pulse is resolved. \\
\underline{edge}: half pulse, i.e. a \emph{rising} or a \emph{falling edge} is observed. In this case the digital pulse is not necessarily resolved as the observation of only one `edge' is already considered as an event. \\ 
\underline{CL}: raised state, i.e. observations where the source is continuously lensed for the complete duration of the survey. \\
\end{enumerate}

The notations ``lens'', ``edge'' and ``CL''
refer to specific detection criteria.  We also introduce the label
``max'' to mean the number of microlensing events between
source, string and observer during the total duration of an experiment
for an ideal detector with infinite time resolution continuously
monitoring the source.\footnote{The expected number of detections per
  source are averaged over the
  random starting times for the microlensing and observation sequences
  and subject to threshold definitions.}
  
As described in \hyperref[app:efficiency]{Appendix~\ref*{app:efficiency}} flux measurements are subject to noise (detector noise, photon shot noise, etc.) and dilution (when the event time-scale is shorter than the exposure time). As part of the detection criteria we select a minimum threshold, ${\delmi}$, below which a flux measurement is considered low or unlensed. Similarly, we identify a maximum downwards fluctuation, ${\delma}$ above which the flux measurement is considered high or lensed. Low, intermediate and high fluxes fall in the ranges $\{1,1+\delmi\}$, $\{1+\delmi,2-\delma\}$ and $\{2-\delma,2\}$, respectively (where $1$ is unlensed). The issues of false dismissal and false detection of microlensing are described in \hyperref[app:efficiency]{Appendix~\ref*{app:efficiency}}. Each detection criterion requires a choice of ${\delmi}$ and ${\delma}$.  Note that in this analysis we do not include any noise sources but we must still select ${\delmi}$ and ${\delma}$ because of the effect of flux averaging.  We regard $\delmi=\delma \le 0.3$ as a practical compromise between minimizing false detections plus false dismissals and maximizing total rates of detection for the current goal of generating estimates. In doing this all sources are regarded as equivalent, however the formalism permits the incorporation of detector noise, Poisson counting statistics, etc. and more sophisticated cuts should ultimately be considered for the analysis of actual survey data. Our specific choices for these inputs are displayed in \autoref{table:SurvSetUp}.

The rate at which geometric alignment occurs is known from $(\Gammalens/{N_\star})/\Nrep$ and Eq.~\ref{eq:Gamma_perStar}. Given that an alignment has taken place we simulate the experiment. To begin the differential lensing rate along the line of sight are sampled, randomly drawing the position according to the probability distribution function. Once the position of the lens is known we infer the time-scale for the duration of the event from the geometry and knowledge of the maximum event duration (which itself depends upon string tension and source distance and $\thetastr$). Next the relative time for the start of observations and the beginning of the time of passage is randomly drawn. This gives two time series: one describes the exposure windows of the experiment and the other the underlying microlensing intervals. From the overlap the expected average flux for each exposure window can be calculated. This vector of fluxes is the elemental result of the simulation. Following this procedure many realizations of the same experiment are created and we apply the detection criteria to count outcomes of different types and average over realizations. Given the known rate of geometric alignments the rate and expected number of each outcome for the experiment can be inferred.

The results are summarised as an efficiency $\epsilon$ such that the expected number of detections according to a given criterion and for a given survey of duration $\Tobs$ is $N_{\rm x} = \epsilon_{\rm x} \, \Gammalens \, \Tobs $, where the subscript ${\rm x}$ stands for the different event definitions: {\rm lens}, {\rm edge}, {\rm CL} or {\rm max}. The expected number of detections per source over the course of the experiment is
\begin{equation}
n_{\rm x} = N_{\rm x}/N_\star  \,= \epsilon_{\rm x} \, \Gammalens \, \Tobs/N_\star  \, ,  
\end{equation}
where we use $n_{\rm x}$ for number per source and $N_{\rm x}$ for total number for all sources. The efficiency accounts for all non-ideal effects: the exposure and microlensing windows may not overlap, the duration of the experiment and the time of passage differ and the observed fluxes are averages over exposure time. Here, $\epsilon_{\rm x}$ is a function of all the experimental parameters including $\Tobs$ and the string tension. For details, see \hyperref[app:efficiency]{Appendix~\ref*{app:efficiency}} and \hyperref[app:efficiencyResults]{Appendix~\ref*{app:efficiencyResults}}.

We also calculate statistical quantities from the flux vectors (without applying any specific detection criterion for events). These results are used to determine the average number of experiments per source and the average number of measurements per source that manifest various general properties. For example, we find the numbers for experiments that report any mix of both low and intermediate values.

%
\begin{figure}
     \includegraphics[trim= 40.0 40.0 40.0 40.0, clip, width=7.7cm]{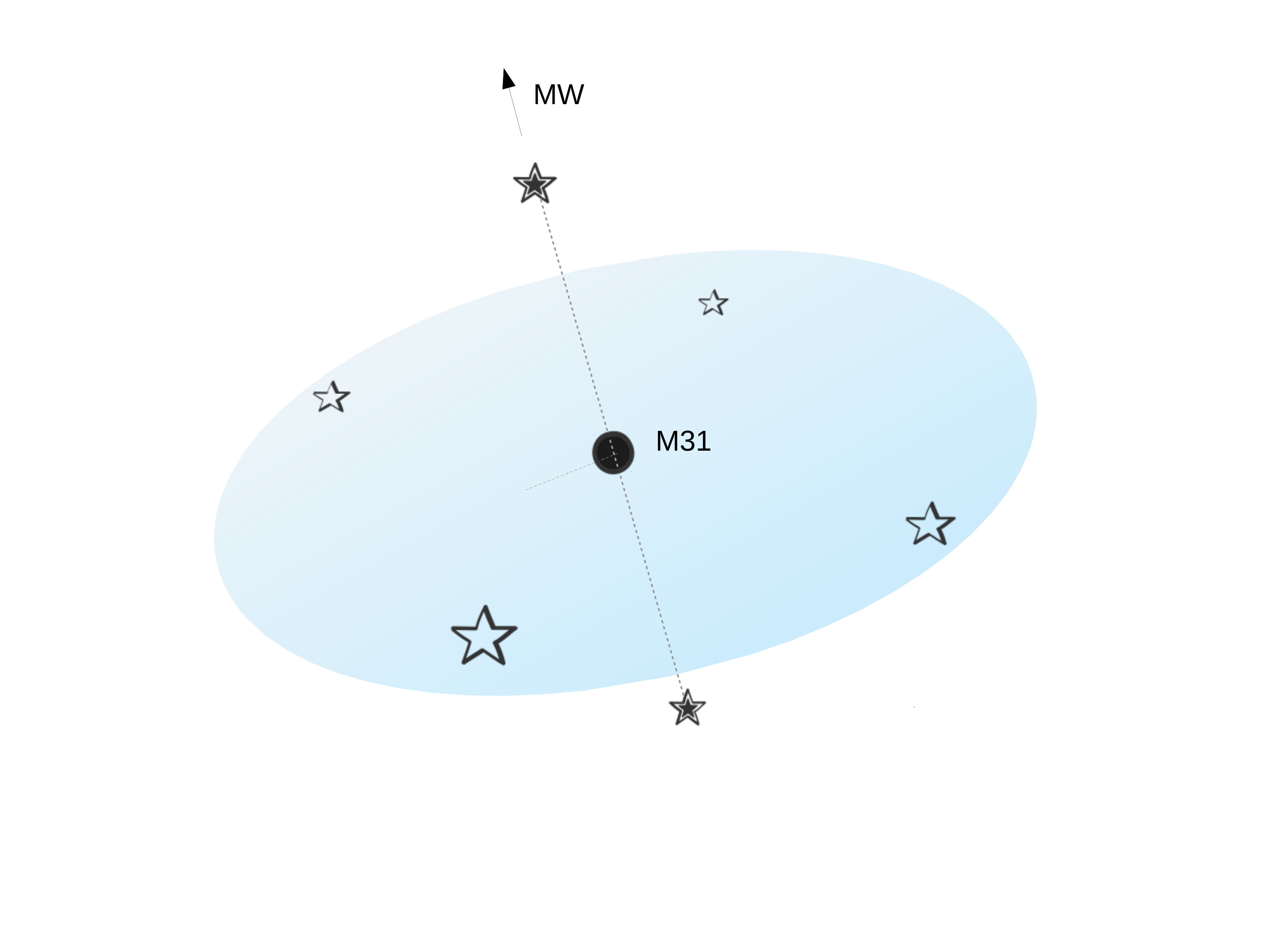}
     \vspace{0.05cm}
 \caption{Positions of six M31 sources (stars equidistant from the centre denoted by the black dot). Four lie in a plane perpendicular to the MW line of sight, one is in front of and one behind M31's centre. We label sources in front of the centre $s_{\rm front}$, behind the centre $s_{\rm back}$ and those in the perpendicular plane $s_{\rm plane}$. \label{fig:analog_pos}}
\end{figure}
%

%
\section{Results} \label{sec:Results}
%

\subsection{Expected Number of Lensing Events for Stars in M31} \label{sec:Res_M31}

\begin{figure*}
 \begin{minipage}{0.8\linewidth}
 \centering
  \includegraphics[clip, width=\hsize]{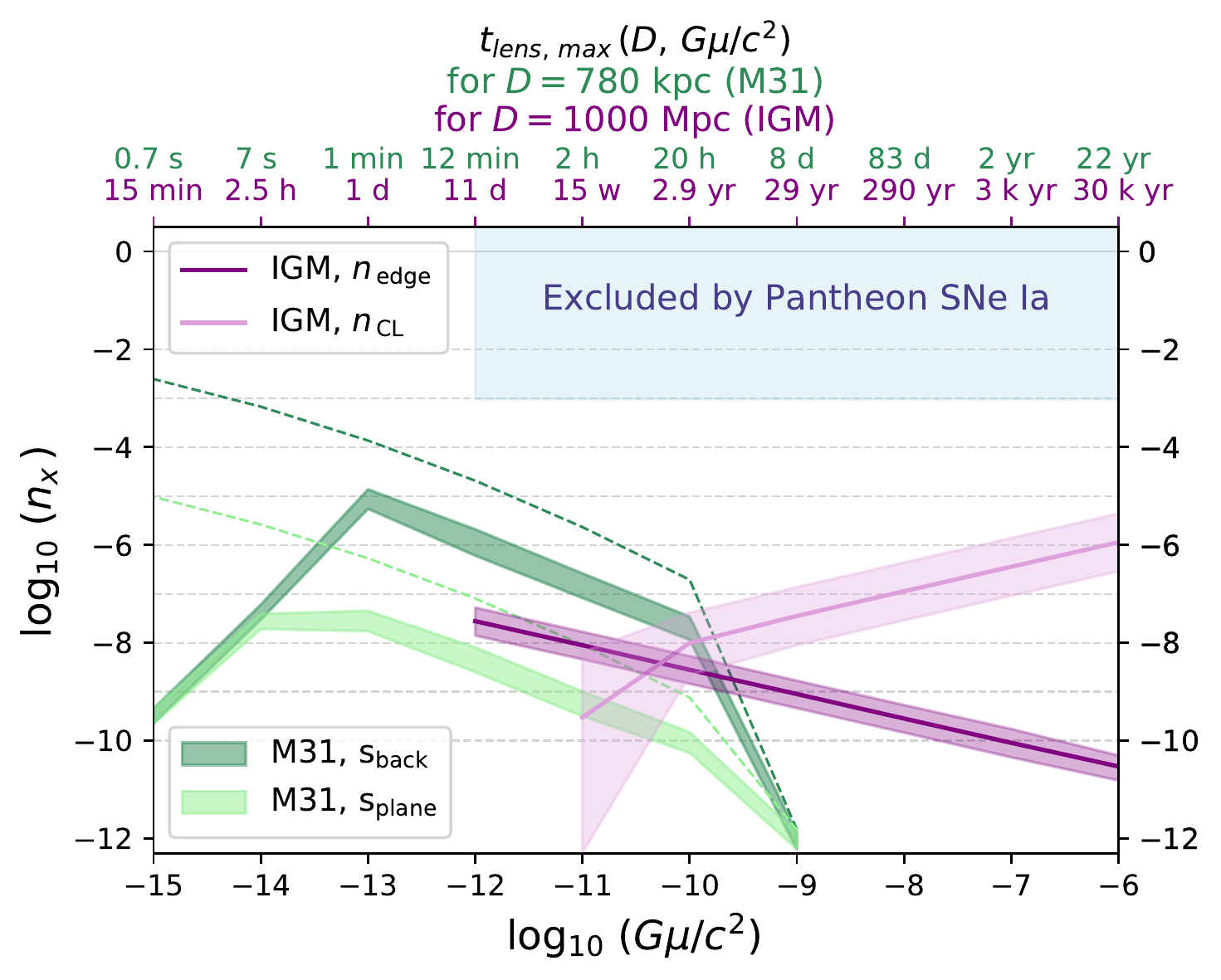}
    \end{minipage}
       \caption{Expected number of detected microlensing events per source as a function of string tension and source position. Both cases, lensing of a star in M31 (\emph{green}) and the lensing of a distant source by cosmic superstrings in the IGM (\emph{purple}) are shown. The upper x-axis indicates the maximum lensing times of one microlensing event typical for a source $8$ kpc from the centre of M31 at distance $\rsource=780$ kpc (\emph{green, upper row}), and for a source at $\rsource= 1000$ Mpc as an example for a source lensed by a string in the IGM (\emph{purple, lower row}). The total duration of the surveys are $\Tobs = 1$ yr for stars in M31 and $\Tobs = 3$ months for the IGM case. The full observation strategies for the surveys for the M31 and IGM observations can be found in \autoref{table:SurvSetUp}. \newline
    \textbf{M31:} The lower edge of the bands indicate the expected number of fully resolved microlensing events, $\nlens$, the upper edges are the event numbers for the detection of a rising or falling edge, $\nedge$. We show the expected number of events for a source located $8$ kpc behind the centre of M31 (\emph{dark green, upper band}) and a source in the plane perpendicular to the line of sight (\emph{light green, lower band}). The \emph{dashed} lines indicates $\nmax$, the expected number of individual microlensing events that take place during the observation period for each source location respectively. This is the maximum number one could detect with an idealised survey, i.e. an instrument of infinite time resolution and constant monitoring of the source during the entire survey period $\Tobs$. 
    Note that the results for a source in front of the centre of M31 almost completely overlap those of a source in the plane of M31; therefore we omitted these bands for clarity.  \newline
    \textbf{IGM:} The expected number of detections of a rising or falling edge, $\nedge$ (\emph{dark purple}), and for the observation of a continuously lensed object, $\ncl$ (\emph{light purple}), are shown. The edges of the bands indicate the results for a source at distance $\rsource=500$ Mpc (\emph{lower edge}), and $\rsource= 2000$ Mpc (\emph{upper edge}). The central line gives the expected numbers for an object at distance $\rsource=1000$ Mpc. These results apply over the whole range of string velocities in the IGM. The blue shaded region marks the region which can be excluded from the null observations of factor two microlensed SNe~Ia in the Pantheon data set \citep{Scolnic:2017caz}.} \label{fig:CombinedRates} 
  
 \end{figure*}

We report results for a prototypical experiment searching for microlensing of stars in M31 by cosmic superstrings with simulation determined network parameters and ${\cal G} = 10^2$. The total survey duration is assumed to be 1 year with 2 hours of observing per night and with individual exposure times of $1$ s. The results for the range of string tensions $10^{-15} < G \mu/c^2< 10^{-9}$ and for various source locations in M31 are presented. \autoref{fig:CombinedRates} summarises the expected number of events per source according to various detection criteria and source positions. 

In particular, we show the expected number of fully resolved events, $\nlens$, and of detecting only one edge, $\nedge$, for a source at $8$ kpc distance from the centre of M31 for the detection criterion $\delmi=\delma=0.3$. The comparison of these expected detections for the assumed survey (see \autoref{table:SurvSetUp}) with the number of events an idealized experiment could detect, $\nmax$, gives a notion of the survey efficiency.

As is evident from \autoref{fig:CombinedRates} the expected number of lens and edge detections for a particular source reaches a maximum for tension $\tension \sim 10^{-13}$. There are two effects at play: (1) the efficiency of detection is a strong function of string tension because $\tlens$ varies with tension. When it becomes less than the exposure time in the prototypical survey, $1$ s, the average flux is diluted and difficult to recognize as lensed. The variation in efficiency with tension can be inferred by comparing the expected number of detections ($\nlens$ for full; $\nedge$ for half pulses) to the expected number in an idealized experiment, $\nmax$. The prototypical survey captures almost all lensing events at $\tension = 10^{-9}$ and less that $0.1$\% at $\tension = 10^{-15}$. On the other hand, (2) the number densities of loops in the halo decreases as string tension rises. The lensing of a source has a lower rate, albeit longer duration, at higher tensions. Together these two effects create the peak in \autoref{fig:CombinedRates} in the expected number of detections.

There is a considerable rate enhancement for sources located behind the centre of M31. These are the most promising targets for tensions $10^{-13} \lesssim \tension \lesssim 10^{-10}$. The number density of string loops tracks the density of the dark matter halo and is large in the centre of the galaxy. A comparison of the results for a source behind the centre of M31 and in the plane shows a gain in number of detections of up to two orders of magnitude. This enhancement diminishes for $\tension < 10^{-13}$ because the strings in the centre of the galaxy are naturally located close to the source, giving rise to very short lensing durations (see \autoref{sec:time-scales}). For tensions smaller than $\tension \sim  10^{-14} $ these short events can not be resolved with an exposure time of $1$ s. The enhancement disappears for $\tension > 10^{-9}$ because loop clustering is ineffective.

In \autoref{fig:CombinedRates} the expected number of microlensing detections of a source behind M31 is calculated for a path with $100$ pc minimum impact parameter. Raising the impact parameter to $1$ kpc lowers the log of the expected number of detections by $\sim 1.2$ for $10^{-13.5} < \tension < 10^{-9.5}$, a relatively small change on the log-log plot (see \hyperref[app:coresize]{Appendix~\ref*{app:coresize}} for numbers). The expected number of detections for sources behind the centre is maximized for a dark matter cusp and is significantly enhanced whether there is a core or a cusp profile at M31's centre.
%
%

We explored the influence of different detection criteria on the number of detected events. The results discussed so far allow a maximum upwards fluctuation of $30\%$ in flux to be considered as unlensed, whereas a measurement has to be larger than $1.7 \times$ the unlensed flux in order to be classified as lensed (i.e. $\delmi=\delma=0.3$). Loosening this bound increases the expected event detections especially for small string tensions, where the lensing time-scales are comparable to or shorter than the exposure time. Using $\delmi=\delma=0.4$ ($\delmi=\delma=0.5$) leads to an enhancement of a factor of $\sim 3$ ($\sim 5$) for $\tension = 10^{-15}$. Loosened detection criteria do not substantially increase the expected number of events but do compromise the power to distinguish the distinct signature of microlensing events by cosmic superstring from noise and from more usual astrophysical source variations. Actual surveys must treat the cutoffs more carefully but our assumptions should suffice for an estimate.

We also studied sources at $2$ kpc from the galactic centre of Andromeda. The differences in expected numbers of detections are factors of a few ($\sim 0.5 - 3$, depending on string tension and source position). Again, considerations like the detailed 3D source distribution will be important for analysing the limits derived from actual surveys but will not greatly effect our estimates here.

Based on current theoretical understanding the string enhancement used in these calculations, ${\cal G} = 10^2$, might be $10^2$ larger or smaller, scaling all rates in \autoref{fig:CombinedRates}.

\subsubsection*{Exclusion by Null Observations}

The duration of a microlensing event $\tlens$ decreases with decreasing string tension. Any survey with exposure times $\tobs  > \tlens$  cannot hope to measure a factor of two flux enhancement caused by a cosmic superstring. Nevertheless, partial overlap of the lensing event and the exposure window will produce an upward excursion in flux. The absence of such excursions may be used to constrain or exclude the existence of strings owing to the lack of observed lensing events. 

There are many possible metrics that can be used for rejection of strings models. Here we consider one statistic based on the presence or absence of an extreme upwards fluctuation of size $\delmi$. Using simulation we compute the expected number of {\it experiments} that see at least one fluctuation $\ge \delmi$ at any time over the entire course of observations of a single star.
We define the expected number of such experiments measuring at least one enhancement when observing one star as $\nav(>\delmi)$. This quantity is determined from the simulations, finding that in all applications $\nav(>\delmi)<<1$ holds. The probability of seeing no fluctuation larger than size $\delmi$ among $N_\star$ independent stars is given by
\begin{equation}
    P_0 = (1 - \nav(>\delmi))^{N_\star} \,. 
\end{equation}
If the model produces excursions $>\delmi$ with near certain probability $1-P_0$ (i.e. small $P_0$) and if the observations detect no such fluctuations then the model can be rejected, taking $1-P_0$ as a qualitative measure of the confidence of rejection.

Fixing $\delmi$ the number of stars necessary to reject a particular model can be inferred. With the definition 
\begin{equation}
{\cal N}=1/\nav(>\delmi)    
\end{equation}
the number of stars needed in the survey to reject a certain string tension is $N_\star = {\cal N} \log(1/P_0)$ for any choice of $P_0$. The results are displayed in terms of ${\cal N}$ which is independent of the confidence level. The actual number of stars is easily derived.

\begin{figure*}
 \begin{minipage}{0.8\linewidth}
 \centering
  \includegraphics[clip, width=\hsize]{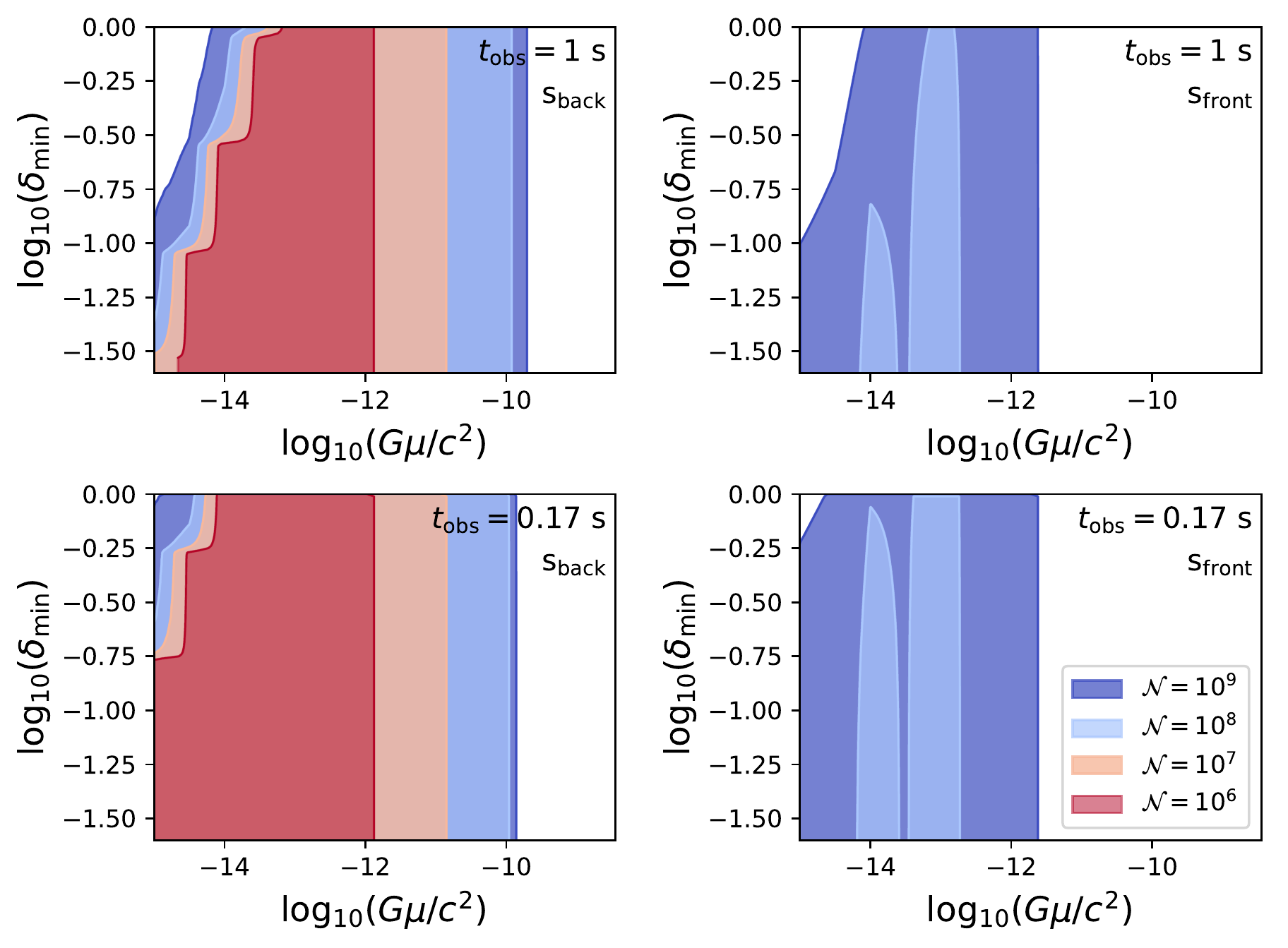}
    \end{minipage}
       \caption{The power of the null detection method to reject string models as a function of ${\cal N}$ and the excursion threshold $\delmi$ for two surveys with $\tobs = 1$ s (\emph{upper panels}) and $0.17$ s (\emph{lower panels}) exposures, $2$ hours per night for $1$ year. The \lp s are for stars behind the centre of M31; the \rp s are for stars in front of M31. The coloured regions indicate which string tensions may be ruled out if observations do not see fluctuations of size $\> \delmi$ in ${\cal N} \log(1/P_0)$ stars. E.g. the \emph{top left} panel shows that at a fixed $\delmi=0.05$ (or $\log_{10} \delmi=-1.3$) the testable tension range is approximately $10^{-14.6} < \tension < 10^{-11.9}$ with ${\cal N}=10^6$.} \label{fig:NullMethod} 
 \end{figure*}
%
%

%
\begin{figure*}
\begin{minipage}{0.45\linewidth}
     \centering
     \includegraphics[trim= 0 0 0 0 0, clip, width=7.9cm]{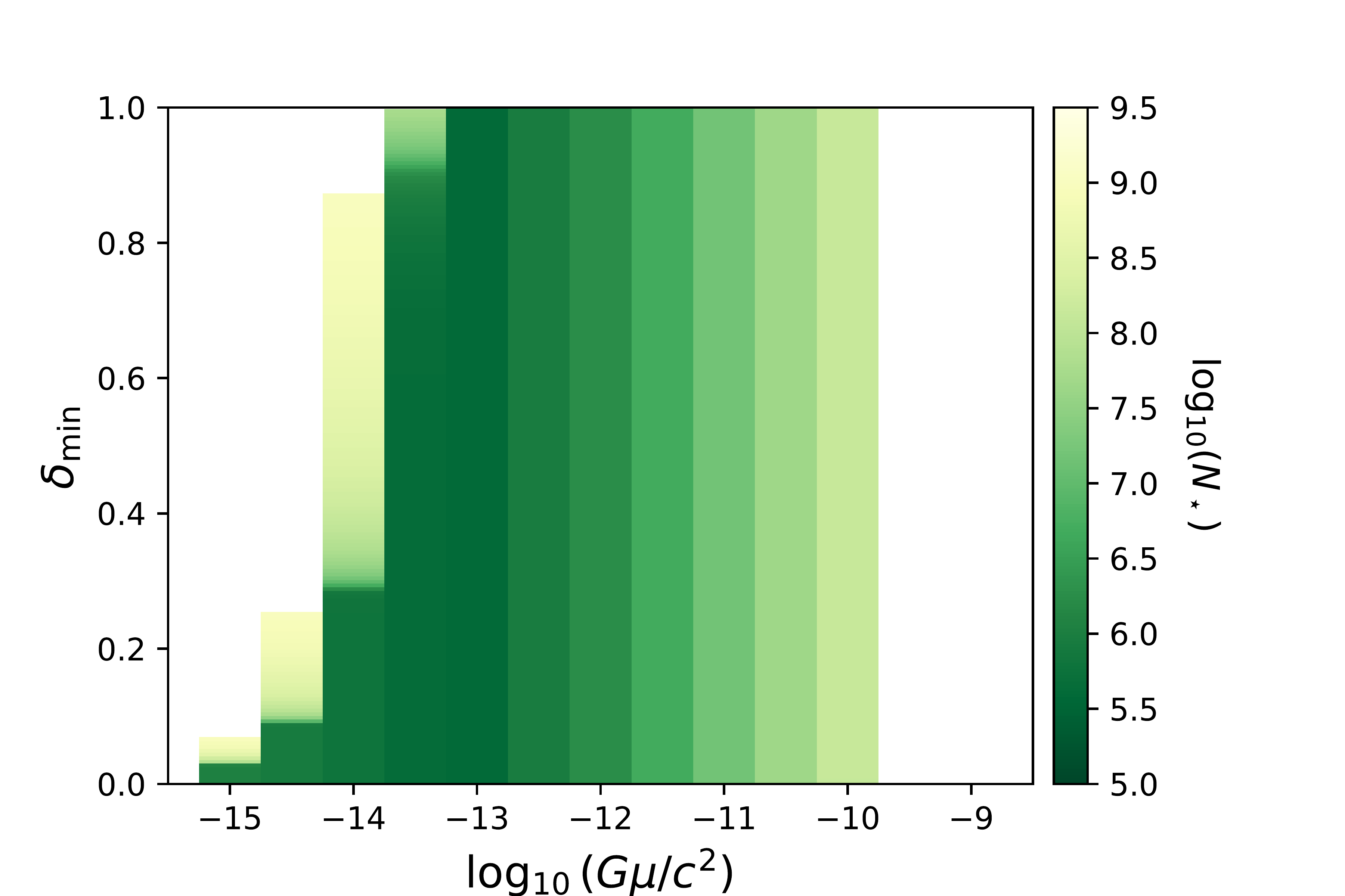}
     \vspace{0.05cm}
    \end{minipage}
    \begin{minipage}{0.45\linewidth}
      \centering\includegraphics[trim= 0.0 0.0 0 0.0 0, clip, width=7.9cm]{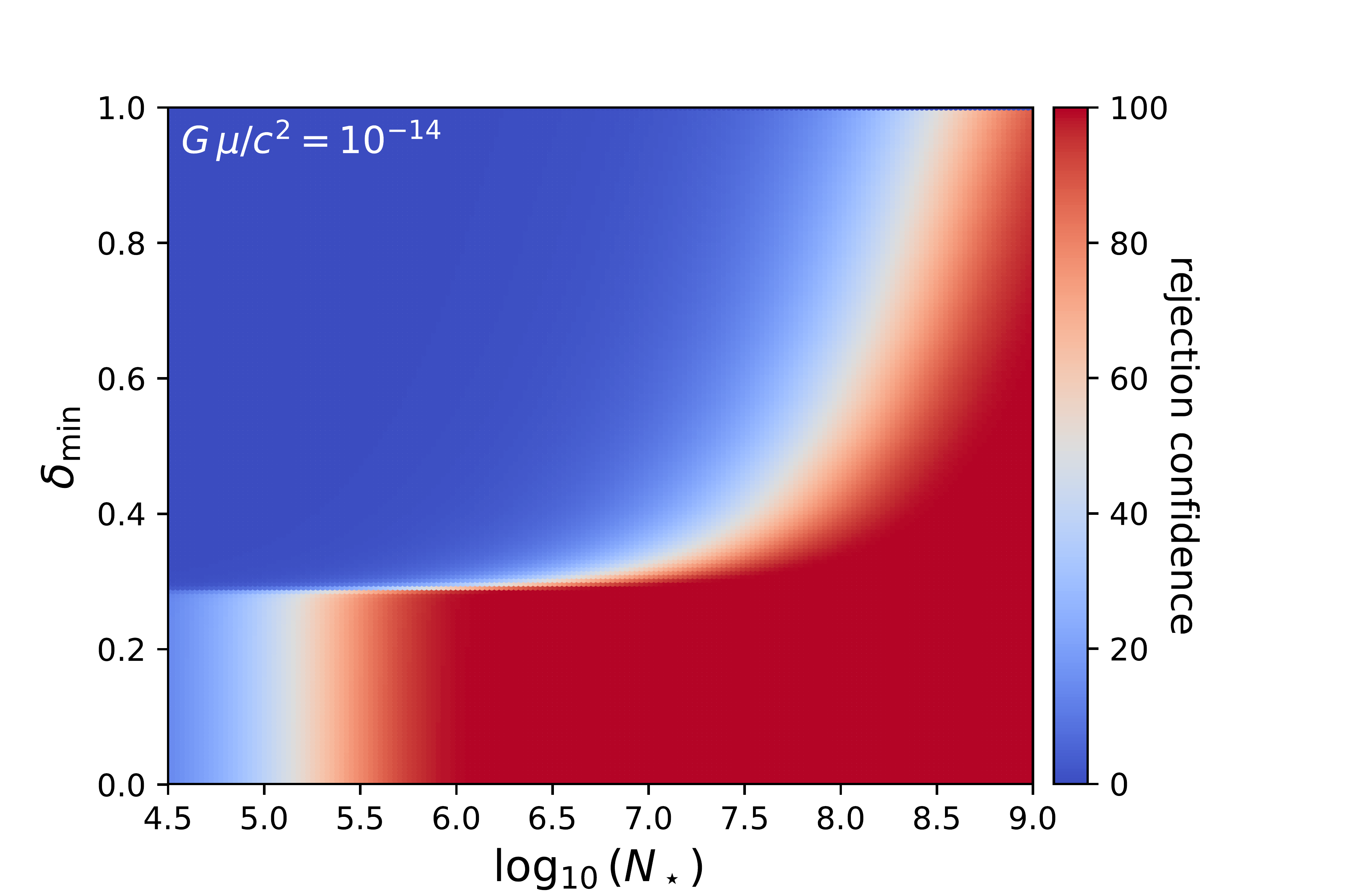}
     \hspace{0.05cm}
    \end{minipage}
 \caption{The power of the null detection method to reject string models as a function of the number of observed stars $N_\star$, and the excursion threshold $\delmi$. The \emph{left panel} shows how many stars behind the centre of M31 would need to be monitored to be able to reject the existence of cosmic superstrings of a given string tension with $95\%$ confidence when no upwards fluctuation of a given $\delmi$ is observed. For string tensions $\tension \gtrsim 10^{-13}$ the lensing durations are longer than the exposure time ($\tobs = 1$ s) such that the factor of two enhancement can always be observed. Therefore the number of stars for a given rejection confidence is independent of the chosen threshold $\delmi$. String tensions $\gtrsim 10^{-9}$ can not be constrained with this method as -- owing to their low number density -- the necessary statistics can not be reached with the observation of up to $10^{9.5}$ stars.  
The \emph{right panel} shows the rejection confidence in the $N_\star - \delmi$ plane for string tension $\tension = 10^{-14}$. For this tension the lensing events partially overlap with the observation window so that the choice of $\delmi$ is crucial for the rejection confidence.  \label{fig:NullMethod2} }
\end{figure*}
%

\autoref{fig:NullMethod} displays the tension ranges that can be ruled out with a survey with given ${\cal N}$. For example, if $\delmi=0.1$ is fixed then the considered prototypical M31 survey ($1$ s exposures, $2$ hours per night, for $1$ year) with ${\cal N}=10^6$ has the power to rule out models with tension $10^{-14.5} < \tension < 10^{-12}$ and ${\cal G}=10^2$ by looking at stars behind the centre (\emph{top left} of \autoref{fig:NullMethod}). Increasing ${\cal N}$ and/or decreasing the threshold enlarges the range that can be probed. (At $99$\% confidence, the red shading is a survey of size $N_\star=4.6 {\cal N} =  4.6 \times 10^6$.) Likewise, ${\cal N}=10^7$ covers $10^{-14.7}  < \tension <10^{-11}$. The \emph{top right} panel is for stars in front of M31 for which much larger numbers are needed to achieve similar constraints.

The exposure time is a critical parameter for the power of the null detection method. The \emph{lower panels} in \autoref{fig:NullMethod} show the same information for a survey that samples at $6$ Hz (and all other survey properties are the same). Smaller tensions become accessible: at $\delmi=0.5$ the lower end of the range extends to $10^{-14.9}<\tension$ and at $\delmi=0.05$ it extends to $10^{-15} < \tension$. Below $\tension \sim 10^{-15}$ the finite size effects of the stars begin to smear the fluxes and the point source approximation is inadequate. The upper limits in the $1$ s and $1/6$ s cadence surveys are essentially identical because the probability of lensing sets the needed sample size and exposure times are short compared to microlensing durations.

The \emph{left panel} in \autoref{fig:NullMethod2} displays the estimated number of stars behind M31 needed to reject at 95\% confidence as a function of string tension and threshold. When the exposure time is short compared to microlensing time the results are insensitive to the choice of $\delmi$ (equivalent to why the right edge of the shadings are vertical in \autoref{fig:NullMethod}). When the exposure time is comparable to or less than the microlensing time the choice of $\delmi$ has a significant impact on the power to reject as displayed in \emph{right panel} in \autoref{fig:NullMethod2}.

In addition to consideration of the relative size of exposure time and microlensing duration the choice of the threshold must ensure that all non-string sources of variation are small. It should be well above that set by known noise sources (photon counting statistics, detector noise, etc.) and astrophysical sources (stellar flares, instabilities, etc.). If the threshold is too low non-string sources will pollute the observations while if it is too high the expected number of fluctuations from models will be too small. In both cases the ability to reject is compromised. 

These examples show that even when the expected number of detected microlensing events is vanishingly small the strings produce potentially observable fluctuations. The lack of observed flux excursions allows an experiment to confidently exclude the existence of superstrings of a given string tension. This general method can be extended to arbitrarily low string tensions but the breakdown of the point source lensing approximation must be taken into account. When the source has a finite size the lensing is diluted and the character of the fluctuations is modified.

\subsection{Expected Lensing Rates of Distant Sources from Cosmic Superstrings in the IGM} \label{sec:Res_IGM}

We consider an experiment searching for microlensing by surveys similar to current SNe~Ia observation programs. Here, we assumed a source is monitored for a period of 3 months, five times a night with each exposure time $30$ s. \autoref{fig:CombinedRates} presents the expected number of events per SNe~Ia where the events are either edge detections or continuously lensed sources, and the thresholds are set by $\delmi = \delma = 0.1$.

The figure shows tension range $10^{-12} < \tension < 10^{-6}$, in which SNe~Ia can still be approximated as point-like sources. The results are shown for source distances of $0.5$, $1$ and $2$ Gpc. The possibility of observing continuous lensing depends on the a priori knowledge of the source flux, something that is possible for SNe~Ia but not for sources with unknown brightness, as e.g. the stars in M31.

The static probability of lensing $\Plens$ (Eq.~\ref{eq:Plens}) dominates the number of sources transiently lensed during the course of the experiment (Eq.~\ref{eq:Rlens}) for $\tension \, \gta \, 10^{-10}$. The scaling of $\Plens$ and $\Rlens$ is $\mu^{1/2}$ and $\mu^{-1/2}$, respectively. This explains the slopes of the two lines.\footnote{The exact values for $\nedge$, $\ncl$, and some intermediate results leading to the numbers shown in \autoref{fig:CombinedRates} can be found in \autoref{table:IGM_edge} and \autoref{table:IGM_cl} in \hyperref[app:efficiencyResults]{Appendix~\ref*{app:efficiencyResults}}.} 
For $\tension \gtrsim 10^{-10}$ the lensing time-scales are several years and the probability of observing a continuously lensed object exceeds that of observing any transition. For $10^{-12} \lesssim \, \tension < \, 10^{-10}$ the time-scales are between days and weeks and the reverse is true. We omit $\nlens$ which is comparable to $\nedge$ and $\nmax$ for $\tension \lta 10^{-11}$. For  $\tension \gta 10^{-10}$ we have $\nlens << \nedge \sim \nmax$. The results for $\nedge$ are independent of the loop center of mass velocity since the product of $\Nrep$ and the rate at which loop alignment occurs is nearly constant. More generally experiments with $\Tobs < \tlens$ are independent of the center of mass velocity and well described by a static lensing probability.

The applied detection criteria have thresholds for up- and downwards fluctuations $\delmi = \delma = 0.1$. Imposing less stringent constraints leads only to minor changes ($ < 0.1 \% $) in the expected number of detections because the exposure times are small compared to the duration of the microlensing event. Hence, a partial overlap of the microlensing event and an observation window is very unlikely. 

A single survey in principle limits both edge and continuous detections but \autoref{fig:CombinedRates} makes it clear that millions of SNe~Ia at $\sim \! 1$ Gpc distances are needed to even begin to constrain strings in the tension range plotted. If we take the accessible range not currently excluded to be $10^{-12} \lta \tension \lta 10^{-8}$ then $\sim \! 10 - 100$ million SNe~Ia at distances $\sim \! 1$ Gpc are needed. Such sample sizes are beyond reach today even with the forthcoming Large Synoptic Survey Telescope (LSST). Therefore, the prospect of constraining the parameter space of cosmic superstrings via this sort of experiment is not very hopeful.

\subsubsection*{Limits from Observations}

Null results from current data sets can be used to set constraints on the string loop density. Even though they are not competitive with bounds from e.g. CMB or gravitational wave experiments they provide an independent constraint. We use the 1048 Pantheon SNe~Ia \citep{Scolnic:2017caz} and simulate the expected event rates for this data set by dividing the SNe~Ia in redshift bins of width $z=0.1$.\footnote{The redshifts are converted into light travel distances by assuming a $\Lambda$CDM cosmology with $H_0 = 67.4$, $\Omega_M = 0.315$ and $\Omega_\Lambda = 0.685$ consistent with the Planck 2018 results \citep{Aghanim:2018eyx}.} The expected number of lensed SNe~Ia in the Pantheon data set is given by
\begin{equation}
    N_{\rm Pantheon} = \sum_i w(z_i) \, n_{\rm x}(z_i) = \sum_i  w(z_i) \, \epsilon_{\rm x}(z_i)  \frac{\Gammalens(z_i)}{N_\star} \Tobs \, ,
\end{equation}
where the index i sums over the different redshift bins, the weight $w(z_i)$ quantifies the number of SNe in one redshift bin, and $n_{\rm x}$ is the number of expected events per source. The latter is given by the product of the efficiency of the survey to detect an microlensing event, $ \epsilon$, and the rate of lensing of a given source, $\Gammalens / N_\star $ (see Eq.~\ref{eq:Gamma_perStar}), the total observation time of the object (here: 3 months). We take $n_{\rm x}(z_i)$ to be the maximum of $n_{\rm edge} (z_i)$ and $n_{\rm CL}(z_i)$ in each redshift bin. Note that $N_{\rm Pantheon}$ depends upon string tension via the detection efficiency and modeling of the string loop distribution.

Visual inspection of the Pantheon light curves finds no anomalous events at the $\delmi=\delma=0.1$ level,\footnote{The choice of $\delta$ makes little difference ($<0.1$\%) in the estimate of $N_{\rm Pantheon}$ and there are no enhanced events (at the 10\% level) in the Pantheon data set.} i.e. no unexpected increase of flux by a factor of at least $\sim 1.9$ has been observed in a SNe~Ia spectrum. One way to interpret this result is that the number of expected events $N_{\rm Pantheon}$ gives a rough upper bound on the string density at each tension. The IGM's string density is the density for SSSUIP scaled by ${\cal G}$ (Eq.~\ref{eq:dn_dlnl_hom}) which accounts for several factors that raise the numbers of superstrings. That value is unknown and we have adopted the fiducial value ${\cal G}=10^2$ up until now. We can recast the null result as an upper bound on ${\cal G}$
\begin{equation}
{\cal G} < \frac{10^2}{N_{\rm Pantheon}} 
\end{equation}
which is larger than the fiducial value. This gives an empirical upper limit for the true IGM distribution
\begin{equation}
    \dndlogl_{\rm IGM}< \left( \frac{10^2}{N_{\rm Pantheon}} \right) \dndlogl_{\rm base}
\end{equation}
where $\dndloglinline_{\rm base}$ continues to be given by Eq.~\ref{eq:dn_dlnl}. Taking the typical string length today to be $l_g$ (i.e. setting $x=1$ in Eq.~\ref{eq:dn_dlnl}) and adopting the simulation determined network parameters for $f$, $\Gamma$ and $\alpha$ leads to an explicit limit on the string density
\begin{equation} 
 \dndlogl_{\rm IGM} \lesssim \, \frac{2.0 \times 10^{-5}}{ N_{\rm Pantheon} \, \mu_{-13}^{3/2}}\,  \, {\rm kpc}^{-3} \, .
\end{equation}
The explicit numerical constraints for different string tensions are shown in \autoref{table:Npantheon}.

%
\begin{table}
\caption{The expected number of microlensing detections (column 2) in the Pantheon data set compromising 1048 SNe~Ia \citep{Scolnic:2017caz} given as a function of string tension (column 1). The corresponding upper limits on the number density of cosmic superstrings, $(\dd \, n / \dd \, {\rm ln} \, l)_{\, \rm hom} = {\cal G} \, (\dd \, n / \dd \, {\rm ln} \, l)_{\, \rm base}$, and on the parameter $\cal G$ in Eq.~\ref{eq:dn_dlnl_hom} appear in last two columns.} 
\label{table:Npantheon}     
\centering 
\begin{tabular}{c|ccc} 
\hline \hline
$\tension$ & $N_{\rm Pantheon}$ & ${\cal G} \, \left( \frac{\dd \, n}{\dd \, {\rm ln}\, l} \right)_{\rm base}$ [kpc$^{-3}$] & ${\cal G}$      \\ \hline   
$ 10^{-6}$ & $1.76 \times 10^{-3}$ & $< \, 3.59 \times 10^{-13}$ & $< \, 5.67 \times 10^{4}$ \\
$ 10^{-7}$ & $5.56 \times 10^{-4}$ & $< \, 3.60 \times 10^{-11}$ & $< \, 1.80 \times 10^{5}$ \\
$ 10^{-8}$ & $1.76 \times 10^{-4}$ & $< \, 3.60 \times 10^{-9}$ & $< \, 5.70 \times 10^{5}$ \\
$ 10^{-9}$ & $1.10 \times 10^{-4}$ & $< \, 1.81 \times 10^{-7}$ & $< \, 9.07 \times 10^{5}$ \\
$ 10^{-10}$ & $1.10 \times 10^{-4}$ & $< \, 5.74 \times 10^{-6}$ & $< \, 9.07 \times 10^{5}$ \\
$ 10^{-11}$ & $1.62 \times 10^{-5}$ & $< \, 1.24 \times 10^{-3}$ & $< \, 6.19 \times 10^{6}$ \\
$ 10^{-12}$ & $9.45 \times 10^{-6}$ & $< \, 6.69 \times 10^{-2}$ & $< \, 1.06 \times 10^{7}$ \\
\hline   \hline        
\end{tabular}
\end{table}
%

%
\section{Conclusions} \label{sec:concl}
%

In this work we study the feasibility of detecting cosmic superstring loops by gravitational microlensing utilising two types of extragalactic sources, stars in our galactic neighbour, Andromeda (M31), and distant Type Ia supernovae (`standard candles').  The microlensing signal is a repeating, achromatic factor of two flux enhancement with sharp rising and trailing edges. In the zoo of transient astrophysical signals string microlensing is distinctive and unlikely to be confused with more prosaic sources. 

In principle, any survey making repetitive flux measurements has some capacity to detect and/or constrain the abundance and tension of superstring loops. Practically speaking microlensing searches probe the low redshift universe. In this respect they are complementary to and independent of other search methods that rely on string imprints on the CMB and a background of gravitational wave emission. Moreover, as we demonstrate any microlensing survey depends crucially on the cadence of observations. 

In this context, we have analysed the capabilities of two prototypical microlensing searches using a detailed description of the underlying string loop distribution while simulating the lensing and the observations in detail.

The two scenarios are lensing of stars in M31 and of SNe~Ia. Together these two studies can potentially detect microlensing over a huge range $10^{-15} < \tension < 10^{-6}$. The unobstructed view of the whole M31 galaxy allows for lines of sight passing through central regions with high dark matter density. This is beneficial for the case of string loops of low tension which cluster with dark matter, increasing the probability of lensing background sources. We construct a model of the MW and M31 dark matter haloes and the string density along the line of sight and simulate the microlensing of stars in M31. We find that the microlensing rate for low tensions is substantial but detection is challenging because the intrinsic time-scales of the expected microlensing events are short, decreasing detection efficiency. 

Lensing of distant sources like SNe~Ia by strings in the intergalactic medium benefits from long path-lengths. String clustering is irrelevant for the IGM lines of sight and, in this sense, SNe~Ia studies are agnostic for low versus high tension strings (the rate of lensing decreases with tension but the instantaneous probability increases). Longer intrinsic time-scales, proportional to distance, make the time of exposure a less critical factor than it is for M31. As tension increases the duration of even a single microlensing event exceeds the experiment's total duration and the source is continuously lensed. In this limit only a standard candle with anomalous flux can reveal the lensing.

The full model of source and detection has many elements. We briefly mention some of the assumptions that have been made in the analysis before presenting the conclusions. The model assumes:
\begin{itemize}
    \item The density of loops is based on a network scaling solution in $\Lambda$CDM cosmology where a fixed fraction $f=0.2$ of the network that is excised by intercommutation goes to form large loops with size $\alpha=0.1$ times the horizon. The scaling solution assumes that the loops decay by gravitational wave losses only ($\Gamma=50$). This is consistent with the most recent studies of the string network.
    \item The primary theoretical parameters to be inferred are the string tension $\mu$ and the enhancement of superstrings with respect to SSSUIP is ${\cal G}$. In brane inflation models both are related to the compactification of the bulk space.
    \item The range of tension of interest is extremely large $10^{-32} < \tension < 10^{-6}$ according to theory and observations. Only a part of that range can be addressed with any one experiment. We focus on assessing the capability of experiments to detect strings with $10^{-15} < \tension < 10^{-6}$ where the lower bound is related to the point source approximation for microlensing and the upper range to the GUT scale.
    \item The range of ${\cal G}$ is at least $1 \le {\cal G} < 10^4$. We take as a working value ${\cal G} = 10^2$ but the full range should be thought of as plausible. 
\end{itemize}

We made a number of simplifying assumptions enumerated in \hyperref[app:simplifications]{Appendix~\ref*{app:simplifications}} for simulating two prototypical surveys. Briefly, these involve crudely sampling the distribution of the stars in M31, making assumptions about the dark matter distribution of M31, taking the low redshift limit for lensing of SN~Ia, utilizing simplified string loop dynamics, not fully sampling all nuisance variables, restricting to high signal-to-noise flux measurements, presuming high spatial resolution to avoid source blending, restricting to point-like sources (no finite size source effects, no unresolved binaries), and ignoring obscuration effects.

To assess the potential of observational programs \textbf{targeting M31}, we simulated a hypothetical survey with 1 second cadence and 0.01 s readout time, taking data for one year, for two hours per night.  Our main findings are:
\begin{itemize}
\item Existing observations of stars in neighbouring galaxies do not yet constrain the tension in cosmic superstring models since the exposure times and cadences are too long to be sensitive to the typical time-scales of the most numerous string microlensing events. 

\item A sample of $\sim 10^5$ stars is needed to begin to constrain the model at $ \tension \sim 10^{-13}$ (see \autoref{fig:CombinedRates} for how this changes with source location and tension).

\item A sample of $\sim 10^6$ stars will generate one microlensing event in the model for string tensions in the range $10^{-13.5} < \tension < 10^{-11.5}$ (see \autoref{fig:CombinedRates}); the constrained range increases with the number of stars.

\item Conversely, string models with $10^{-14} < \tension < 10^{-12}$ may be excluded at 99\% confidence by failing to detect 10\% fluctuations in $4 \times 10^6$ stars laying behind the centre of M31.

\item The cadence of time sampling sets the lower tension limit that the null method probes. At $6$ Hz, the lower limit is less than $\sim 10^{-15}$ (same confidence, number of stars and fluctuation threshold).

\end{itemize}

Since time-sampling is a critical aspect of sensitivity of microlensing surveys targeting nearby galaxies, it is important to point out the technological trends. The readout times of CCDs cameras have serious limitations for the study of transient phenomena lasting a few seconds or less but CMOS detectors have the potential to open up this new frontier in optical astronomy. CMOS technology allows cadence observations of several Hz \citep{2016SPIE.9915E..04J} and are being used to instrument wide-field cameras aimed at fast time-domain astronomy, e.g.,  the 1.3-meter Transneptunian Automated Occultation Survey \cite[TAOS II;][]{2017DPS....4921614L}, the 0.55 meter Weizmann Fast Astronomical Survey Telescope \citep[W-FAST;][]{2017AAS...22915506N}, and the Tomo-e Gozen wide-field CMOS camera for the Kiso 1.0-m telescope \citep{2016SPIE.9908E..3PS}. The improvements promise to permit overlapping coverage of the entire tension range shown range $10^{-15} < \tension < 10^{-11.5}$. While resolved observations of a large sample of stars in the crowded bulge region remain a considerable practical challenge they are not entirely unfeasible \citep{Han:1995gv}.

The \textbf{search for cosmic superstrings with SNe~Ia} complements the search with stars in nearby galaxies in the sense that SNe~Ia surveys are more sensitive to higher string tensions. Since the expected brightness of the sources are known it is possible to identify objects which are continuously lensed by a string during the whole observation period. Generally, higher string tensions give rise to longer lensing events making the occurrence of a continuously lensed source more likely. Such sources will show up as outliers in the Hubble diagram. We find:
\begin{itemize}
\item Detecting superstrings from the model with tensions $10^{-12} < \tension < 10^{-6} $ will require many SN~Ia. We estimate $10^7 - 10^8$ are needed for a reasonable probability of success.
\item SNe~Ia observations constrain the number density of cosmic superstrings of a given tension by setting upper limits on the string theory enhancement ${\cal G}$, see \autoref{table:Npantheon}.
\end{itemize}
While forthcoming surveys are unlikely to yield sufficient SNe~Ia, the possibility that an event will be discovered by chance remains since the standard observational strategy of monitoring SNe~Ia does not have to be changed. 

In conclusion, year-long experiments targeting stars in nearby galaxies provide the possibility to probe a wide range of string tensions
for the string model. The capability of these experiments to detect a microlensing event sourced by cosmic superstrings increases with the number of stars monitored and with decreasing exposure and readout time. A detection of a SNe~Ia lensed by a cosmic superstring is not likely in view of the size of upcoming data sets. However, typical survey strategies are sensitive to these events leaving open the possibility for a detection by chance. The absence of flux fluctuations may be used to provide additional constraints on the string models, especially when individual microlensing events cannot be resolved in time.

\section*{acknowledgements}
We are very grateful to David Marsh, Liam McAllister, Sterl Phinney, Maxim Pshirkov, Bo Sundberg and Henry Tye for helpful comments on the manuscript.
D.F.C. acknowledges the John Templeton Foundation New Frontiers Program Grant No. 37426 (University of Chicago) and FP050136-B (Cornell University) and
NSF Grant No. 1417132. A.G. acknowledges support from the Swedish Research Council and the Swedish National Space Agency. J.J.R. acknowledges support by Katherine Freese through a grant from the Swedish Research Council (Contract No. 638-2013-8993).

%

\appendix

%
\section{Efficiency Simulation} \label{app:efficiency}
%

This appendix serves to outline the efficiency calculation. A loop's motion across the sky is broken up into two components, one that is the loop's centre of mass and the other that is its internal motion. Let the loop have the invariant size $l$ at distance $\rstring$ moving with characteristic centre of mass velocity $\vcom$. The characteristic internal velocity is much larger than the centre of mass motion for cosmologically old loops, i.e. $\vosc >> \vcom$. The angular area per unit time swept out by the centre of mass motion is $\Delta {\dot \Omega} \equiv (1/\rstring^2)\, \dd \, A_{\rm com}/\dd \, t$ where $\dd \, A_{\rm com}/ \dd \, t$ is the rate at which the (period averaged) projected area of the loop traverses the sky. The characteristic scale is $\dd \, A_{\rm com}/\dd \, t \sim l \vcom$. In any instance it depends upon geometric factors describing the specific orientation of the loop centre of mass velocity, the orientation of the loop with respect to the line of sight and the oscillation figure generated by the loop. An alignment occurs when the (period averaged) projected area of the loop crosses a background star. For the rate estimates $\dd \, A_{\rm com}/\dd \, t$ has been averaged over orientations.

For a background stellar density $\Sigma_*$ (number per angular area) the rate of generation of alignments per string is $\Rgeo= \Sigma_* \Delta {\dot \Omega}$. For a foreground string density $\Sigma_{\rm str}$ in terms of number per angular area the rate of generation of alignments per angular area of the sky is $\Sigma_{\rm str} \, \Rgeo$.

During a time interval $T$ the expected number of new alignments per string is $\Rgeo T$.  If $\Rgeo T << 1$, as is true for all the examples we consider, there are effectively two cases (1) no alignment with probability approximately $1-\Rgeo T$, and (2) one alignment with probability approximately $\Rgeo T$.

When an alignment occurs there are approximately $\Nrep$ microlensing events that take place during the entire time it takes the loop to traverse the line of sight to the star. The loop crosses the background source 2, 4, 6 ...
times per period. We conservatively assume two crossings per period, creating two subsequences. Each subsequence of microlensing epochs is exactly periodic. We assume that the two subsequences have a random phase offset in time. The microlensing rate during the alignment is $\Nrep/\tpass$ where $\tpass \sim l/\vcom$. Each microlensing event has characteristic length $\tlens << \tpass$. The expected rate of occurrence of microlensing events without regard to alignment is $\Rmicro = \Rgeo \Nrep$ (no alignment contributes zero events and one alignment contributes $\Nrep$ events). During the time $T$ the expected number of microlensing events is $\Rmicro T$.

The number of stars (strings) in a small angular area $\Delta \Omega$ is $N_\star = \Sigma_* \Delta \Omega$ ($N_{\rm str} = \Sigma_{\rm str} \Delta \Omega$). The intrinsic physical microlensing rate (number per time for all the stars in the survey) is $\Gammalens = \Sigma_{\rm str} \Rmicro \Delta \Omega = N_{\rm str} \Rmicro$. The microlensing rate per star $\Gammalens/ N_\star = (\Sigma_{\rm str}/\Sigma_*) \Rmicro$.

An ideal experiments makes individual measurements with exposure times less than $\tlens$ and without gaps and for time $T >> \tlens$. When microlensing occurs the typical microlensing event is fully contained within the span of $T$. The signature is unambiguous, a factor of two, i.e. observable as a digital step function in flux. This is the ideal circumstance and the detection criterion is simple.

The ideal situation not realised when
\begin{itemize}
\item the monitoring is not continuous; microlensing that occurs between exposures may be completely missed,
\item the microlensing time-scale is less than an individual exposure time; the factor of two distinction between lensed and unlensed flux is lost,
\item the microlensing time-scale is comparable to or longer than $T$; one or both of the low/high transitions of the digital step will be missed.
\end{itemize}
These complications lower the efficiency of any real experiment and it is important to understand what combination of string tension and time-scales a given experiment is sensitive to. Instead of requiring the resolution of a full microlensing event, i.e. observing both low-to-high and high-to-low transitions, it is possible to loosen the detection criterion by hunting for the edges of a flux enhancement event. Each choice of detection criterion entails an efficiency calculation.

The efficiency for an arbitrary detection criterion is analysed as follows:

The experimental strategy (epochs for exposures and times of exposure for the entire set of observations) and the string tension are stipulated. The source distance is chosen according to the experiment. In the case of M31, all stars are at an approximately fixed distance ($\rsource=780$ kpc); in the case of supernovae the distances are set individually. The baseline string loop's distribution is assumed and we model the string loop distribution along the line of sight. For M31 the dark matter haloes are scaled to give the inhomogeneous distribution (this includes the tension-dependent clustering); for the IGM we use the homogeneous distribution (see \autoref{sec:cosmo}).

The string loop's location along the line of sight is drawn according to the total rate at which loops lens the source (averaged over orientation effects and integrated over the loop size distribution). In an ideal case the rate, conditional on the line of sight distance would need to be sampled to infer $l$, $\thetastr$ and $\vcomperp$. In this work we use typical values and assume the loop has the typical size $l = l_g$, the string-line of sight orientation $\thetastr=\pi/4$, and the centre of mass velocity perpendicular to the line of sight $\vcom = 150$ km/s for string in the MW or in the Andromenda galaxy. For cosmic superstrings residing in the IGM we use $\vcom = 600$ km/s. These fixed choices set the time-scale for the duration of the maximum microlensing event and the crossing time as well as the number of microlensing events per alignment. The CDF for microlensing times scales with the maximum microlensing time and is the same for all choices along the line of sight. Consequently, the choice of distance implies the microlensing time (see \autoref{fig:time-scales}).

Once the distance is selected, there is a rigidly defined flux amplification sequence with an unknown starting point in time. The experimental strategy is also a rigidly defined sequence in time. The relative time offset for the start of the flux amplification and of the experiment is randomly selected from the total time interval $T$ during which any amplification is possible. 

The total time $T$ corresponds to the sum of the crossing time $\tpass$ and the total duration of the experiment $\Tobs$ (note, the duration is not the exposure time but instead the time from the start of the first exposure to the end of the last exposure). The ideal expected number of alignments is $\Rgeo T$ and the ideal expected number of microlensing events is $\Rmicro T = \Rgeo \Nrep \, T$ where $T=\tpass+\Tobs$.

A vector of expected average fluxes is reported with a component $f_i$ for each successive exposure. When an exposure window does not overlap a microlensing event, the result is 1; when the window falls fully within an event, the result is 2; when there is partial overlap, the result is the weighted average of 1 and 2. (There is also a small but finite possibility for higher fluxes if both amplification subsequences overlap during one exposure window.)

Various detection criteria can be evaluated with the flux vector. For example, if noise is ignored then the criterion closest to the ideal digital pulse is a discrete subsequence of flux measures  $\dots 1$,$2 \dots 2$,$1 \dots $ . Let $N_x$ be the number of occurrences of sequences like this within one flux vector that satisfy detection criterion "x". We repeat the random sampling for the string geometry and the phases of the subsequences and evaluate $\bar N_x$, which is the average number of detected events during time $T$. 

The ideal number of microlensing events per experiment of length $\Tobs$ is $\Rmicro \Tobs= \Rgeo \Tobs \Nrep$. We define the efficiency for the experiment of length $T$ in detecting events $\epsilon_x  = (T/\Tobs)({\bar N_x}/\Nrep)$ so that the expected number of detected events during the experiment of length $\Tobs$ is $\Rmicro \Tobs \epsilon = \Rgeo T {\bar N_x}$.

When defining one event differently, e.g. as the rise of a flux ($1 \dots 1$,$2 \dots$) or the fall ($2 \dots 2$,$1 \dots$), different detection efficiencies $\epsilon_x$ with different expected event numbers $N_x$ can be defined (this example corresponds to the definition of $\Nedge$).  

When noise is present for each component of the flux vector $f_i$ a ``measured'' flux $d_i$ can be drawn from a Poisson distribution based on the expected number.  Even when an individual flux is 1 or 2 (no partial overlap of the microlensing and observation windows) the measured flux is no longer an integer. 

With a good sample of flux vectors ${\vec d}$ various statistics of interest can be formulated. Scanning through the history of a single star rising and falling edges are defined by the criteria
\begin{itemize}
\item If $d_i < 1 + \delma$ and $d_{i+1} > 2-\delmi$ there is a rising edge,
\item If $d_i > 2 - \delmi$ and $d_{i+1} < 1 + \delma$ there is a falling edge.
\end{itemize}
where $\delma$ and $\delmi$ are cutoffs. The number of rising edges $N_+$ and falling edges $N_-$ give a (lower) estimate for the number of microlensing events $N = (N_+ + N_-)/2$ and mean number $\Nav$.

In the example above the choices of $\delma$ and $\delmi$ are arbitrary and make a significant difference, defining the threshold for what constitutes a largely unlensed or largely lensed star. As one rough choice assume $N_\star$ unlensed stars are observed each $\Nobs$ times such that $\Nobs N_\star = 10^{10}$. A fairly simple recipe is to demand less than one false detection of flux doubling in the $10^{10}$ measurements and less than $1\%$ chance of false dismissal of a fully lensed star observed one time. Taking the flux errors to be Gaussian $\propto e^{-(d-1)^2/2\sigma^2}$ for the unlensed stars, suppressing false detection implies that $d < \dma$ where $\dma=1+\delma= 1 + 6.36 \sigma$ should be identified as unlensed (i.e., the cumulative probability for $d>d_{max}$ is less than $10^{-10}$). Likewise taking the flux errors to be Gaussian $\propto e^{-(d-2)^2/2\sigma^2}$ for the lensed stars, suppressing false dismissal implies $d > \dmi$ where $\dmi=2 - \delmi = 2-2.33 \sigma$ should be identified as lensed (i.e., the cumulative probability for $d<\dmi$ is less than $1\%$). To summarise, the acceptable range is a function of $\sigma$. If $\sigma$ corresponds to 5\% of the flux then $\delma=0.32$ and $\delmi=0.12$ satisfy these criteria. We also applied these criteria for all the produced efficiency estimates; the expected number of events would be decreased by a factor of $2-3$ for stars in M31 and sub-percent level in the case of SNe being lensed by strings in the IGM.


%
\section{Simplifications in treatment} \label{app:simplifications}
%
In addition to the simplifications made for describing the string loop distribution we have made a number of simplifications in the drawing the lenses, sources and treatment of the microlensing. Many of these are physical approximations but others neglect known physical effects.
\begin{itemize}
    \item (Approximation to M31) We analyse selected source positions for stars. We do not analyse the full distribution. The range of sources subsumes the bulk that a survey of the central regions of M31 will record.
    \item (Approximation to M31) We approximate the dark matter distribution in M31 by a cusp profile and impose a minimum impact parameter $100$ pc.
    \item (Low redshifts for SN~Ia) The lensing treatment is suitable for loops at small redshift ranges $0 \le z < 2$; it ignores the rate variations associated with the loop density's redshift dependence; it ignores and the cosmological effects on lensing geometry. This produces conservative lensing rates.
    \item (Separate halo and IGM effects) We ignore lensing contributions from the MW halo and the SN~Ia halo. This produces conservative lensing rates.
    \item (Simplified loop dynamics) We assume and simulate two exactly periodic microlensing sequences when a loop passes over a background source. This ignores that the source is lensed differently as the source traverses the sky, breaking the exact periodicity of the loop oscillations. We have assumed exact periodicity in the signal. In addition, the number of microlensing sequences can increase above two. Our restriction to two subsequences is a conservative assumption.
    \item (Restricted sampling) In the simulation many nuisance variables have been fixed at typical values rather than sampled. These include the relative velocity of the observer, centre of mass of the loop and source; the orientation of the loop with respect to the line of sight; the velocity of the loop perpendicular to the line of sight at the time of microlensing. 
    \item (High S/N detections) No photon noise and no detector noise is included.
    \item (High spatial resolution) Background light in the source pixel is ignored.
    \item (Point-like sources) Complex lensing signals (finite source size, multiple unresolved stars) are ignored.
    \item (Unobstructed seeing) Extinction effects are not included.
\end{itemize}

%
\section{Numerical Results from Efficiency Simulation} \label{app:efficiencyResults}
%

\begin{table*}
\caption{Lensing rates per star $\Gammalens/ N_\star$, simulated detection efficiency $\epslens$, and number of events per source when observing M31 for $\Tobs = 1$ yr for 2 hours per night with 1 s exposure time and 0.01 s readout time. The event number $\nlens$ is given by the product $\Gammalens/N_\star \times \Tobs \times \epslens$. The first three columns indicate the according quantities for a source located behind the centre of M31, the last three for a source located in the plane perpendicular to the line of sight.}            
\label{table:M31}     
\centering                       
\begin{tabular}{c|ccc |ccc }    
\hline \hline
 & \textbf{M31} & Analog:& source$_{\rm back}$ &      Analog:& source$_{\rm plane}$  &  \\ \hline
log$_{10}(G \mu / c^2)$ &  $\Gammalens/N_\star \times \Tobs$  & $\epslens$  &  $\nlens$ &  $\Gammalens/N_\star \times \Tobs$  & $\epslens$  &  $\nlens$  \\

\hline  
-09 & $1.52 \times 10^{-12 }$ & $3.92 \times 10^{-1 }$ & $5.98 \times 10^{-13 }$ & $1.51 \times 10^{-12 }$ & $4.06 \times 10^{-1 }$ & $6.12 \times 10^{-13}$ \\
-10 & $1.95 \times 10^{-7 }$ & $5.92 \times 10^{-2 }$ & $1.16 \times 10^{-8 }$ & $7.67 \times 10^{-10 }$ & $7.44 \times 10^{-2 }$ & $5.71 \times 10^{-11}$ \\
-11 & $2.33 \times 10^{-6 }$ & $3.56 \times 10^{-2 }$ & $8.29 \times 10^{-8 }$ & $9.14 \times 10^{-9 }$ & $3.47 \times 10^{-2 }$ & $3.17 \times 10^{-10}$ \\
-12 & $2.07 \times 10^{-5 }$ & $2.94 \times 10^{-2 }$ & $6.07 \times 10^{-7 }$ & $8.11 \times 10^{-8 }$ & $3.12 \times 10^{-2 }$ & $2.53 \times 10^{-9}$ \\
-13 & $1.36 \times 10^{-4 }$ & $4.01 \times 10^{-2 }$ & $5.47 \times 10^{-6 }$ & $5.36 \times 10^{-7 }$ & $3.24 \times 10^{-2 }$ & $1.74 \times 10^{-8}$ \\
-14 & $6.72 \times 10^{-4 }$ & $4.44 \times 10^{-5 }$ & $2.98 \times 10^{-8 }$ & $2.64 \times 10^{-6 }$ & $7.30 \times 10^{-3 }$ & $1.93 \times 10^{-8}$ \\
-15 & $2.46 \times 10^{-3 }$ & $9.27 \times 10^{-8 }$ & $2.28 \times 10^{-10 }$ & $9.66 \times 10^{-6 }$ & $2.32 \times 10^{-5 }$ & $2.24 \times 10^{-10}$ \\
\hline   \hline                         
\end{tabular}
\end{table*}

\begin{table*}
\caption{Lensing rates per star $\Gammalens/ N_\star$, simulated detection efficiency $\epsedge$, and expected number of detected microlensing events on a cosmic superstring in the IGM per source for  $\Tobs = 3$ months. In this table the number of detections, $\nedge = \Gammalens/ N_\star \times \Tobs \times \epsedge$, is given by requiring to detect either the flux increase or the flux decrease of a microlensing event, i.e. seeing only one edge.  We assume 5 observations per night with an exposure time of 30 s. The first three columns indicate the according quantities for a source at $\rsource=500$ Mpc distance, the last three for a distance of $\rsource=1$ Gpc.}    
\label{table:IGM_edge}     
\centering                         
\begin{tabular}{c|ccc|ccc }      
\hline \hline
 & \textbf{IGM} & distance & $\rsource = 500$ Mpc &  & distance & $\rsource = 1000$ Mpc      \\ \hline
log$_{10}(G \mu / c^2)$ &  $\Gammalens/N_\star \times \Tobs$  & $\epsedge$  &  $\nedge$ &  $\Gammalens/N_\star \times \Tobs$  & $\epsedge$  &  $\nedge$  \\

\hline   
-06 & $7.66 \times 10^{-12 }$ & $2.02 $ & $1.54 \times 10^{-11 }$ & $1.53 \times 10^{-11 }$ & $1.96 $ & $2.99 \times 10^{-11}$ \\
-07 & $2.42 \times 10^{-11 }$ & $1.90 $ & $4.59 \times 10^{-11 }$ & $4.84 \times 10^{-11 }$ & $1.88 $ & $9.10 \times 10^{-11}$ \\
-08 & $7.66 \times 10^{-11 }$ & $1.91 $ & $1.47 \times 10^{-10 }$ & $1.53 \times 10^{-10 }$ & $1.86 $ & $2.85 \times 10^{-10}$ \\
-09 & $2.42 \times 10^{-10 }$ & $1.92 $ & $4.64 \times 10^{-10 }$ & $4.84 \times 10^{-10 }$ & $1.87 $ & $9.04 \times 10^{-10}$ \\
-10 & $7.66 \times 10^{-10 }$ & $1.91 $ & $1.46 \times 10^{-9 }$ & $1.53 \times 10^{-9 }$ & $1.84 $ & $2.82 \times 10^{-9}$ \\
-11 & $2.42 \times 10^{-9 }$ & $1.91 $ & $4.62 \times 10^{-9 }$ & $4.84 \times 10^{-9 }$ & $1.85 $ & $8.97 \times 10^{-9}$ \\
-12 & $7.66 \times 10^{-9 }$ & $1.84 $ & $1.41 \times 10^{-8 }$ & $1.53 \times 10^{-8 }$ & $1.82 $ & $2.79 \times 10^{-8}$ \\
\hline   \hline                                
\end{tabular}
\end{table*}

\begin{table*}
\caption{Same as \autoref{table:IGM_edge} but for the number of detection $\ncl = \Gammalens/ N_\star \times \Tobs \times \epscl$. Here one event is defined by the observation of a SNe~Ia with a flux enhancement of at least $90 \, \%$ during the whole observation time of $\Tobs = 3$ months. We assume 5 observations per night with an exposure time of 30 s.}             
\label{table:IGM_cl}    
\centering                     
\begin{tabular}{c|ccc|ccc }        
\hline \hline
 & \textbf{IGM} & distance & $\rsource = 500$ Mpc &  & distance & $\rsource = 1000$ Mpc      \\ \hline
log$_{10}(G \mu / c^2)$ &  $\Gammalens/N_\star \times \Tobs$  & $\epscl$  &  $\ncl$ &  $\Gammalens/N_\star \times \Tobs$  & $\epscl$  &  $\ncl$  \\

\hline   
-06 & $7.66 \times 10^{-12 }$ & $3.79 \times 10^{+04 }$ & $2.90 \times 10^{-7 }$ & $1.53 \times 10^{-11 }$ & $7.44 \times 10^{4 }$ & $1.14 \times 10^{-6}$ \\
-07 & $2.42 \times 10^{-11 }$ & $3.79 \times 10^{+03 }$ & $9.17 \times 10^{-8 }$ & $4.84 \times 10^{-11 }$ & $7.44 \times 10^{3 }$ & $3.60 \times 10^{-7}$ \\
-08 & $7.66 \times 10^{-11 }$ & $3.78 \times 10^{+02 }$ & $2.89 \times 10^{-8 }$ & $1.53 \times 10^{-10 }$ & $7.44 \times 10^{2 }$ & $1.14 \times 10^{-7}$ \\
-09 & $2.42 \times 10^{-10 }$ & $3.69 \times 10^{+01 }$ & $8.94 \times 10^{-9 }$ & $4.84 \times 10^{-10 }$ & $7.35 \times 10^{1 }$ & $3.56 \times 10^{-8}$ \\
-10 & $7.66 \times 10^{-10 }$ & $2.87 $ & $2.20 \times 10^{-9 }$ & $1.53 \times 10^{-9 }$ & $6.52 $ & $9.99 \times 10^{-9}$ \\
-11 & $2.42 \times 10^{-9 }$ & $2.13 \times 10^{-4 }$ & $5.17 \times 10^{-13 }$ & $4.84 \times 10^{-9 }$ & $6.17 \times 10^{-2 }$ & $2.99 \times 10^{-10}$ \\
\hline   \hline   
\end{tabular}
\end{table*}

We present the exact numbers and results for the lensing rate of a source by a cosmic superstring, $\Gammalens$, the efficiency of the considered survey strategies to detect these events $\epslens$, and the resulting expected number of detected events, $\nlens = \Tobs \epslens \Gammalens / N_\star$. The results for the lensing of a star in M31 are given in \autoref{table:M31}, and of a distant source lensed by a string in the IGM in \autoref{table:IGM_edge} and \autoref{table:IGM_cl} for the event numbers for $\nedge$ and $\ncl$ respectively.

%
\section{Effect of core size of M31} \label{app:coresize}
%

We evaluate the effect of the core size of M31 for events involving sources located directly behind the centre of M31. The core size is $0.1$ kpc in \autoref{fig:CombinedRates} which displays the expected number of events per source $\nlens$ (dark green, upper level) and the maximum number of events per source $\nmax$ (dark dashed green). \autoref{table:coresizes} provides numerical results for both core sizes.

\begin{table*}
  \caption{The expected number of microlensing events per source as a
    function of string tension for the M31 experiment with two
    different halo core sizes, $1$ kpc and $0.1$ kpc. $\nlens$
    describes detected, fully resolved events; $\nmax$ quantifies all
    events occurring during the experiment's 1 year duration.}
\label{table:coresizes}    
\centering                     
\begin{tabular}{c|cc|cc}        
\hline \hline
log$_{10}(G \mu / c^2)$ & log$_{10}\nlens$ ($1$) & log$_{10}\nlens$ ($0.1$) & log$_{10}\nmax$ ($1$) & log$_{10}\nmax$ ($0.1$) \\ \hline
$ -14.5$ & $ -8.1$ & $ -8.$ & $ -4.1$ & $ -2.9 $ \\
$ -14.$ & $ -7.5$ & $ -7.5$ & $ -4.4$ & $ -3.2 $ \\
$ -13.5$ & $ -6.2$ & $ -4.9$ & $ -4.7$ & $ -3.5 $ \\
$ -13.$ & $ -6.5$ & $ -5.3$ & $ -5.1$ & $ -3.9 $ \\
$ -12.5$ & $ -7.$ & $ -5.6$ & $ -5.5$ & $ -4.3 $ \\
$ -12.$ & $ -7.4$ & $ -6.2$ & $ -5.9$ & $ -4.7 $ \\
$ -11.5$ & $ -7.9$ & $ -6.6$ & $ -6.4$ & $ -5.1 $ \\
$ -11.$ & $ -8.4$ & $ -7.1$ & $ -6.9$ & $ -5.6 $ \\
$ -10.5$ & $ -8.8$ & $ -7.6$ & $ -7.4$ & $ -6.2 $ \\
$ -10.$ & $ -9.2$ & $ -7.9$ & $ -8.$ & $ -6.7 $ \\
$ -9.5$ & $ -12.5$ & $ -12.5$ & $ -11.6$ & $ -11.6 $ \\
$ -9.$ & $ -12.3$ & $ -12.3$ & $ -11.8$ & $ -11.8 $ \\
$ -8.5$ & $ -12.1$ & $ -12.1$ & $ -12.1$ & $ -12.1 $ \\
$ -8.$ & $ -12.3$ & $ -12.3$ & $ -12.3$ & $ -12.3 $ \\
$ -7.5$ & $ -12.6$ & $ -12.6$ & $ -12.6$ & $ -12.6 $ \\
$ -7.$ & $ -12.9$ & $ -13.1$ & $ -12.8$ & $ -12.8 $ \\
\hline \hline
\end{tabular}
\end{table*}

%
\section{Definition of Symbols} \label{app:symbols}
%
%
Definitions of symbols according to where in the text they
are first introduced are given in \autoref{table:symbols}, \autoref{table:symbols1} and \autoref{table:symbols2}.

\begin{table*}
\caption{Symbols and definitions for physical quantities}
\label{table:symbols}
\centering
\begin{tabular}{ll}
\hline
\multicolumn{2}{l}{ABSTRACT}\\
\hline
$\mu$ & string tension ($\mu_{x}$ with numerical $x$ stands for the scaled, dimensionless tension $(G \mu/c^2)/10^x$)\\
\hline
\multicolumn{2}{l}{1 INTRODUCTION}\\
\hline
$\Lambda$ & GUT energy scale \\
$E_p$ & Planck energy scale \\
\hline
\multicolumn{2}{l}{2 COSMOLOGY OF SUPERSTRINGS}\\
\hline
SSSUIP & single species of string having unit intercommutation probability \\
$t_{\rm life}$ & proper time for loop to evaporate completely by gravitational wave emission \\
$l$ & invariant (proper) length of a loop \\
$l_g$ & length of a loop that evaporates in a time equal to the age of the Universe \\
$\Gamma$ & dimensionless, loop-dependent factor for the rate of gravitational wave emission ($\Gamma_x$ with numerical $x$ stands for $\Gamma/x$)\\
$\vinter$ & speed of string centre of mass from intercommutation \\
$f$ & fraction of long horizon crossing strings chopped into large loops ($f_{x}$ with numerical $x$ stands for $f/x$) \\
$\alpha$ & size of the large loops relative to the cosmological horizon  ($\alpha_{x}$ with numerical $x$ stands for $\alpha/x$)\\

$\dndl$ & differential number density of string loops per invariant length in scaling solution; subscripts: \\
 & ``base'' - unclustered, SSSUIP \\
 & ``homog'' - unclustered, superstring (multiple species, bound states, reduced intercommutation probability) \\
 & ``inhomog'' - clustered superstring \\

DM & dark matter \\
$t_H$ & age of the Universe\\
$\rho_{\rm DM}$ & spatially varying dark matter density today \\
${\cal E}$ & ratio of spatially varying dark matter density to the average dark matter density throughout the universe today; \\
& dark matter clustering \\
${\cal F}$ & ratio of spatially varying string density to average string density throughout the universe today; \\
& tension-dependent string clustering \\
${\cal G}$ & ratio of homogeneous superstring density to base string density; enhancement of string density for superstrings \\
& compared to SSSUIP \\
${\beta}$ & ${\cal F}/{\cal E}$, the tension-dependent string enhancement from clustering divided by the local dark matter enhancement;\\
& $\beta=1$ when strings cluster like dark matter \\
$\rstring$ ($\rsource$) & line of sight distance from observer to string (observer to source) \\
$\rho_{\rm MW-M31}$ & dark matter density in Milky Way-Andromeda system \\
$r$ & displacement from Galactic centre \\
$A$, $B$, $r_1$, $r_2$ &  model parameters for MW-M31 system \\
\hline
\end{tabular}
\end{table*}

\begin{table*}
\caption{Symbols and definitions for physical quantities 3-7}
\label{table:symbols1}
\centering
\begin{tabular}{ll}
\hline
\multicolumn{2}{l}{3 STRING MICROLENSING}\\
\hline
$\Delta \Theta$ & deficit angle induced by infinite straight string \\
$\left( \delta \phi \right)_{v=0}$ & the angular splitting induced by an infinite straight string at rest with respect to the observer \\
$\delta \phi$ & the angular splitting induced by an infinite straight string \\
$u (u_n)$ & velocity of segment of string (with respect to the observer, along the line of sight) responsible for angular splitting \\
$\thetastr $ & angle of string with respect to line of sight \\
$\tlens$ & duration of a single microlensing event \\
$\Rlens$ & microlensing event rate \\
$\Plens$ & probability for microlensing \\
$\vcom$ ($\vcomperp$) & loop centre of mass velocity (perpendicular to line of sight) \\
$\Nrep$ & number of microlensing events during alignment \\
$\tosc$ & fundamental period of loop \\
$\tpass$ & duration of loop's alignment with background source \\
alignment & epoch when the loop's projected area (time-averaged over $\tosc$) overlaps a background source \\
microlensing event & instance when the observer, string and source are nearly aligned so that multiple paths allow light to reach the observer from the source; \\
& ideally a segment of string passes over a point-like source; the flux, not the angular splitting, is observed \\

\hline
\multicolumn{2}{l}{4 OBSERVATIONAL LIMITS}\\
\hline
$R_\odot$ & radius of the Sun \\
$R_{\rm sh}$ & radius of supernova shell \\
$R_{\rm QSO}$ & physical size of QSO optical emission region \\

\hline
\multicolumn{2}{l}{5 TIME-SCALE OF EVENTS}\\
\hline

$\tlensm$ & the longest duration lensing events when the string is halfway between the observer and source \\
$v_\perp$ & velocity of the string perpendicular to the line of sight \\
CDF & cumulative distribution function \\

\hline
\multicolumn{2}{l}{6 EVENT RATE, EFFICIENCY}\\
\hline

Experiment & record of fluxes according to a given exposure protocol of a set of sources; the key parameters are\\
& number of sources, total time span of the experiment, and number and properties of the exposures \\
$N_{\star}$ & number of stars in a survey cone \\
$\Tobs$ & total duration of survey (from start of first exposure to end of last exposure) \\
$\tobs$ & duration of a single exposure \\
$\Delta \tobs$ & interval between successive exposures in a single night (from end of one exposure to beginning of the next) \\
$\Gammalens$ & intrinsic physical microlensing rate in the survey volume (all sources and strings) \\
$n_{\star}(r)$ & density of stars along the line of sight \\
$< \dd \, A_\perp/ \dd \, t> $ & average area perpendicular to line of sight swept per time by an oscillating string loop \\
$\epsilon_x$ & Efficiency such that $\epsilon_x \Gammalens$ is the
detectable rate for criterion $x$ for a given exposure protocol; $x$ may be \\
& ``lens'' - consecutive up and down transition detected (low-high-low or high-low-high sequence)\\
& ``edge'' - up or down transition detected (low-high or high-low sequence)\\
& ``CL'' - continuously lensed (always high)\\
& ``max'' - lensing detected with infinite time resolution, continuous monitoring over the total duration of the experiment \\
$\delmi$ & low flux (require flux $<1+\delmi$ where $1$ is unlensed) \\
$\delma$ & high flux (require flux $>1+\delma$ where $1$ is unlensed) \\
$N_x$ & expected number of detected microlensing events for criterion $x$ for experiment (for all sources observed) \\
$n_x$ & expected number of detected microlensing events for criterion $x$ per source in experiment \\
$s_x$ & source position with respect to the centre of M31; $x$ is front, back, plane meaning in front of the centre, behind the centre, \\
& in the plane passing through the centre perpendicular to the line of sight, respectively, see \autoref{fig:analog_pos}. \\
\hline
\multicolumn{2}{l}{7 RESULTS}\\
\hline
${\bar n}(>\delta)$, $1/{\cal N}$ & expected number of observation protocols generating at least 1 exposure with flux enhancement $>1+\delta$ for 1 source \\
$P_0$ & probability that experiment (one protocol, $N_\star$ independent sources) has no flux-enhanced exposures in any source \\
$w$ & number of supernova per unit redshift \\
$N_{\rm Pantheon}$ & expected number of lensed SNe Ia in the Pantheon data set  \\

\hline
\end{tabular}
\end{table*}

\begin{table*}
\caption{Symbols and definitions for physical quantities Appendix A}
\label{table:symbols2}
\centering
\begin{tabular}{ll}
\hline
\multicolumn{2}{l}{APPENDIX A}\\
\hline
$\vosc$ & loop oscillatory velocity about centre of mass \\
$\Delta {\dot \Omega}$ & averaged rate angular area is swept out by loop centre of mass motion \\
$\dd \, A_{\rm com}/ \dd \, t$ & average rate at which projected area is swept out by loop centre of mass motion \\
$\Sigma_*$ & surface density of stars; sources per angular area \\
$\Rgeo$ & rate of alignments per string \\
$\Sigma_{\rm str}$ & surface density of string loops; strings per angular area \\
$T$ & time interval \\
$\Rmicro$ & microlensing rate per string \\
$f_i$ & expected flux of i-th exposure based on the protocol \\
$d_i$ & measured flux when $f_i$ is the expected flux \\
$N_x$ & number of detected microlensing events for criterion $x$ for experiment (one realization) \\
${\bar N}_x$ & $N_x$ averaged over realizations of all nuisance parameters for string and source \\
${\epsilon}_x$ & efficiency of experiment for detection for criterion $x$ \\
$N_+ (N_-)$ & number of rising (falling) edges \\
$N$ & number of microlensing events \\

\hline
\end{tabular}
\end{table*}

\eject
\bibliography{refs}

\begin{thebibliography}{}
\makeatletter
\relax
\def\mn@urlcharsother{\let\do\@makeother \do\$\do\&\do\#\do\^\do\_\do\%\do\~}
\def\mn@doi{\begingroup\mn@urlcharsother \@ifnextchar [ {\mn@doi@}
  {\mn@doi@[]}}
\def\mn@doi@[#1]#2{\def\@tempa{#1}\ifx\@tempa\@empty \href
  {http://dx.doi.org/#2} {doi:#2}\else \href {http://dx.doi.org/#2} {#1}\fi
  \endgroup}
\def\mn@eprint#1#2{\mn@eprint@#1:#2::\@nil}
\def\mn@eprint@arXiv#1{\href {http://arxiv.org/abs/#1} {{\tt arXiv:#1}}}
\def\mn@eprint@dblp#1{\href {http://dblp.uni-trier.de/rec/bibtex/#1.xml}
  {dblp:#1}}
\def\mn@eprint@#1:#2:#3:#4\@nil{\def\@tempa {#1}\def\@tempb {#2}\def\@tempc
  {#3}\ifx \@tempc \@empty \let \@tempc \@tempb \let \@tempb \@tempa \fi \ifx
  \@tempb \@empty \def\@tempb {arXiv}\fi \@ifundefined
  {mn@eprint@\@tempb}{\@tempb:\@tempc}{\expandafter \expandafter \csname
  mn@eprint@\@tempb\endcsname \expandafter{\@tempc}}}

\bibitem[\protect\citeauthoryear{Aasi et~al.}{Aasi et~al.}{2014}]{Aasi:2013vna}
Aasi J.,  et~al., 2014, \mn@doi [Phys. Rev. Lett.]
  {10.1103/PhysRevLett.112.131101}, 112, 131101

\bibitem[\protect\citeauthoryear{Abbott et~al.}{Abbott
  et~al.}{2007}]{Abbott:2006vg}
Abbott B.,  et~al., 2007, \mn@doi [Phys. Rev.] {10.1103/PhysRevD.76.082001},
  D76, 082001

\bibitem[\protect\citeauthoryear{Abbott et~al.}{Abbott
  et~al.}{2009a}]{Abbott:2009ws}
Abbott B.~P.,  et~al., 2009a, \mn@doi [Nature] {10.1038/nature08278}, 460, 990

\bibitem[\protect\citeauthoryear{Abbott et~al.}{Abbott
  et~al.}{2009b}]{Abbott:2009rr}
Abbott B.~P.,  et~al., 2009b, \mn@doi [Phys. Rev.]
  {10.1103/PhysRevD.80.062002}, D80, 062002

\bibitem[\protect\citeauthoryear{Abbott et~al.}{Abbott
  et~al.}{2017}]{TheLIGOScientific:2016dpb}
Abbott B.~P.,  et~al., 2017, \mn@doi [Phys. Rev. Lett.]
  {10.1103/PhysRevLett.118.121101, 10.1103/PhysRevLett.119.029901}, 118, 121101

\bibitem[\protect\citeauthoryear{Abbott et~al.}{Abbott
  et~al.}{2018}]{Abbott:2017mem}
Abbott B.,  et~al., 2018, \mn@doi [Phys. Rev.] {10.1103/PhysRevD.97.102002},
  D97, 102002

\bibitem[\protect\citeauthoryear{Ade et~al.}{Ade et~al.}{2014}]{Ade:2013xla}
Ade P. A.~R.,  et~al., 2014, \mn@doi [Astron. Astrophys.]
  {10.1051/0004-6361/201321621}, 571, A25

\bibitem[\protect\citeauthoryear{Aghanim et~al.}{Aghanim
  et~al.}{2018}]{Aghanim:2018eyx}
Aghanim N.,  et~al., 2018

\bibitem[\protect\citeauthoryear{Bahcall, Gramann  \& Cen}{Bahcall
  et~al.}{1994}]{Bahcall:1994mj}
Bahcall N.~A.,  Gramann M.,   Cen R.,  1994, \mn@doi [Astrophys. J.]
  {10.1086/174877}, 436, 23

\bibitem[\protect\citeauthoryear{Battye \& Moss}{Battye \&
  Moss}{2010}]{Battye:2010xz}
Battye R.,  Moss A.,  2010, \mn@doi [Phys. Rev.] {10.1103/PhysRevD.82.023521},
  D82, 023521

\bibitem[\protect\citeauthoryear{Battye, Caldwell  \& Shellard}{Battye
  et~al.}{1997}]{Battye:1997ji}
Battye R.~A.,  Caldwell R.~R.,   Shellard E. P.~S.,  1997, in {Topological
  defects in cosmology}. pp 11--31 (\mn@eprint {arXiv} {astro-ph/9706013})

\bibitem[\protect\citeauthoryear{Baumann \& McAllister}{Baumann \&
  McAllister}{2015}]{Baumann:2014nda}
Baumann D.,  McAllister L.,  2015, {Inflation and String Theory}.
Cambridge Monographs on Mathematical Physics, Cambridge University Press
  (\mn@eprint {arXiv} {1404.2601}), \mn@doi{10.1017/CBO9781316105733}, \url
  {http://www.cambridge.org/mw/academic/subjects/physics/theoretical-physics-and-mathematical-physics/inflation-and-string-theory?format=HB}

\bibitem[\protect\citeauthoryear{Bennett \& Bouchet}{Bennett \&
  Bouchet}{1989}]{Bennett:1989ch}
Bennett D.~P.,  Bouchet F.~R.,  1989, \mn@doi [Phys. Rev. Lett.]
  {10.1103/PhysRevLett.63.1334}, 63, 1334

\bibitem[\protect\citeauthoryear{Bennett et~al.,}{Bennett
  et~al.}{1996}]{Bennett:1996ce}
Bennett C.~L.,  et~al., 1996, \mn@doi [Astrophys. J.] {10.1086/310075}, 464, L1

\bibitem[\protect\citeauthoryear{Bernardeau \& Uzan}{Bernardeau \&
  Uzan}{2001}]{Bernardeau:2000xu}
Bernardeau F.,  Uzan J.-P.,  2001, \mn@doi [Phys. Rev.]
  {10.1103/PhysRevD.63.023005}, D63, 023005

\bibitem[\protect\citeauthoryear{Bevis, Hindmarsh, Kunz  \& Urrestilla}{Bevis
  et~al.}{2007}]{Bevis:2007qz}
Bevis N.,  Hindmarsh M.,  Kunz M.,   Urrestilla J.,  2007, \mn@doi [Phys. Rev.]
  {10.1103/PhysRevD.76.043005}, D76, 043005

\bibitem[\protect\citeauthoryear{Blanco-Pillado \& Olum}{Blanco-Pillado \&
  Olum}{2017}]{Blanco-Pillado:2017oxo}
Blanco-Pillado J.~J.,  Olum K.~D.,  2017, \mn@doi [Phys. Rev.]
  {10.1103/PhysRevD.96.104046}, D96, 104046

\bibitem[\protect\citeauthoryear{Blanco-Pillado, Olum  \&
  Siemens}{Blanco-Pillado et~al.}{2018}]{Blanco-Pillado:2017rnf}
Blanco-Pillado J.~J.,  Olum K.~D.,   Siemens X.,  2018, \mn@doi [Phys. Lett.]
  {10.1016/j.physletb.2018.01.050}, B778, 392

\bibitem[\protect\citeauthoryear{Bloomfield \& Chernoff}{Bloomfield \&
  Chernoff}{2014}]{Bloomfield:2013jka}
Bloomfield J.~K.,  Chernoff D.~F.,  2014, \mn@doi [Phys. Rev.]
  {10.1103/PhysRevD.89.124003}, D89, 124003

\bibitem[\protect\citeauthoryear{Bouchet \& Bennett}{Bouchet \&
  Bennett}{1990}]{Bouchet:1989ck}
Bouchet F.~R.,  Bennett D.~P.,  1990, \mn@doi [Phys. Rev.]
  {10.1103/PhysRevD.41.720}, D41, 720

\bibitem[\protect\citeauthoryear{Caldwell \& Allen}{Caldwell \&
  Allen}{1992}]{Caldwell:1991jj}
Caldwell R.~R.,  Allen B.,  1992, \mn@doi [Phys. Rev.]
  {10.1103/PhysRevD.45.3447}, D45, 3447

\bibitem[\protect\citeauthoryear{Chernoff}{Chernoff}{2009}]{Chernoff:2009tp}
Chernoff D.~F.,  2009, {arXiv:0908.4077}

\bibitem[\protect\citeauthoryear{Chernoff \& Tye}{Chernoff \&
  Tye}{2007}]{Chernoff:2007pd}
Chernoff D.~F.,  Tye S. H.~H.,  2007, {arXiv:0709.1139}

\bibitem[\protect\citeauthoryear{Chernoff \& Tye}{Chernoff \&
  Tye}{2015}]{Chernoff:2014cba}
Chernoff D.~F.,  Tye S. H.~H.,  2015, \mn@doi [Int. J. Mod. Phys.]
  {10.1142/S0218271815300104}, D24, 1530010

\bibitem[\protect\citeauthoryear{Chernoff \& Tye}{Chernoff \&
  Tye}{2018}]{Chernoff:2017fll}
Chernoff D.~F.,  Tye S. H.~H.,  2018, \mn@doi [JCAP]
  {10.1088/1475-7516/2018/05/002}, 1805, 002

\bibitem[\protect\citeauthoryear{Christiansen, Albin, James, Goldman, Maruyama
  \& Smoot}{Christiansen et~al.}{2008}]{Christiansen:2008vi}
Christiansen J.~L.,  Albin E.,  James K.~A.,  Goldman J.,  Maruyama D.,   Smoot
  G.~F.,  2008, \mn@doi [Phys. Rev.] {10.1103/PhysRevD.77.123509}, D77, 123509

\bibitem[\protect\citeauthoryear{Damour \& Vilenkin}{Damour \&
  Vilenkin}{2000}]{Damour:2000wa}
Damour T.,  Vilenkin A.,  2000, \mn@doi [Phys. Rev. Lett.]
  {10.1103/PhysRevLett.85.3761}, 85, 3761

\bibitem[\protect\citeauthoryear{Damour \& Vilenkin}{Damour \&
  Vilenkin}{2001}]{Damour:2001bk}
Damour T.,  Vilenkin A.,  2001, \mn@doi [Phys. Rev.]
  {10.1103/PhysRevD.64.064008}, D64, 064008

\bibitem[\protect\citeauthoryear{Damour \& Vilenkin}{Damour \&
  Vilenkin}{2005}]{Damour:2004kw}
Damour T.,  Vilenkin A.,  2005, \mn@doi [Phys. Rev.]
  {10.1103/PhysRevD.71.063510}, D71, 063510

\bibitem[\protect\citeauthoryear{DePies \& Hogan}{DePies \&
  Hogan}{2007}]{DePies:2007bm}
DePies M.~R.,  Hogan C.~J.,  2007, \mn@doi [Phys. Rev.]
  {10.1103/PhysRevD.75.125006}, D75, 125006

\bibitem[\protect\citeauthoryear{Economou, Harari  \& Sakellariadou}{Economou
  et~al.}{1992}]{Economou:1991bc}
Economou A.,  Harari D.,   Sakellariadou M.,  1992, \mn@doi [Phys. Rev.]
  {10.1103/PhysRevD.45.433}, D45, 433

\bibitem[\protect\citeauthoryear{Fich \& Tremaine}{Fich \&
  Tremaine}{1991}]{Fich:1991ej}
Fich M.,  Tremaine S.,  1991, \mn@doi [Ann. Rev. Astron. Astrophys.]
  {10.1146/annurev.aa.29.090191.002205}, 29, 409

\bibitem[\protect\citeauthoryear{Fraisse}{Fraisse}{2007}]{Fraisse:2006xc}
Fraisse A.~A.,  2007, \mn@doi [JCAP] {10.1088/1475-7516/2007/03/008}, 0703, 008

\bibitem[\protect\citeauthoryear{{Goobar} \& {Leibundgut}}{{Goobar} \&
  {Leibundgut}}{2011}]{2011ARNPS..61..251G}
{Goobar} A.,  {Leibundgut} B.,  2011, \mn@doi [Annual Review of Nuclear and
  Particle Science] {10.1146/annurev-nucl-102010-130434}, \href
  {http://adsabs.harvard.edu/abs/2011ARNPS..61..251G} {61, 251}

\bibitem[\protect\citeauthoryear{{Graham} et~al.,}{{Graham}
  et~al.}{2019}]{2019arXiv190201945G}
{Graham} M.~J.,  et~al., 2019, arXiv:1902.01945, \href
  {http://adsabs.harvard.edu/abs/2019arXiv190201945G} {}

\bibitem[\protect\citeauthoryear{Han}{Han}{1996}]{Han:1995gv}
Han C.,  1996, \mn@doi [Astrophys. J.] {10.1086/178045}, 472, 108

\bibitem[\protect\citeauthoryear{Hogan}{Hogan}{2006}]{Hogan:2006we}
Hogan C.~J.,  2006, \mn@doi [Phys. Rev.] {10.1103/PhysRevD.74.043526}, D74,
  043526

\bibitem[\protect\citeauthoryear{{Hogan} \& {Narayan}}{{Hogan} \&
  {Narayan}}{1984}]{Hogan:1984unknown}
{Hogan} C.,  {Narayan} R.,  1984, \mn@doi [\mnras] {10.1093/mnras/211.3.575},
  \href {http://adsabs.harvard.edu/abs/1984MNRAS.211..575H} {211, 575}

\bibitem[\protect\citeauthoryear{Jenet et~al.,}{Jenet
  et~al.}{2006}]{Jenet:2006sv}
Jenet F.~A.,  et~al., 2006, \mn@doi [Astrophys. J.] {10.1086/508702}, 653, 1571

\bibitem[\protect\citeauthoryear{{Jorden}, {Jerram}, {Jordan}, {Pratlong}  \&
  {Robbins}}{{Jorden} et~al.}{2016}]{2016SPIE.9915E..04J}
{Jorden} P.,  {Jerram} P.,  {Jordan} D.,  {Pratlong} J.,   {Robbins} M.,  2016,
  in High Energy, Optical, and Infrared Detectors for Astronomy VII. p. 991504,
  \mn@doi{10.1117/12.2239429}

\bibitem[\protect\citeauthoryear{Kaspi, Taylor  \& Ryba}{Kaspi
  et~al.}{1994}]{Kaspi:1994hp}
Kaspi V.~M.,  Taylor J.~H.,   Ryba M.~F.,  1994, \mn@doi [Astrophys. J.]
  {10.1086/174280}, 428, 713

\bibitem[\protect\citeauthoryear{Kibble}{Kibble}{1976}]{Kibble:1976sj}
Kibble T. W.~B.,  1976, \mn@doi [J. Phys.] {10.1088/0305-4470/9/8/029}, A9,
  1387

\bibitem[\protect\citeauthoryear{Kogut et~al.}{Kogut
  et~al.}{1993}]{Kogut:1993ag}
Kogut A.,  et~al., 1993, \mn@doi [Astrophys. J.] {10.1086/173453}, 419, 1

\bibitem[\protect\citeauthoryear{{Kuijken}, {Siemens}  \&
  {Vachaspati}}{{Kuijken} et~al.}{2008}]{2008MNRAS.384..161K}
{Kuijken} K.,  {Siemens} X.,   {Vachaspati} T.,  2008, \mn@doi [\mnras]
  {10.1111/j.1365-2966.2007.12663.x}, \href
  {http://adsabs.harvard.edu/abs/2008MNRAS.384..161K} {384, 161}

\bibitem[\protect\citeauthoryear{{Lehner} et~al.,}{{Lehner}
  et~al.}{2017}]{2017DPS....4921614L}
{Lehner} M.,  et~al., 2017, in AAS/Division for Planetary Sciences Meeting
  Abstracts \#49. p. 216.14

\bibitem[\protect\citeauthoryear{{Nir}, {Ofek}, {Ben-Ami}, {Manulis},
  {Gal-Yam}, {Diner}  \& {Rappaport}}{{Nir} et~al.}{2017}]{2017AAS...22915506N}
{Nir} G.,  {Ofek} E.~O.,  {Ben-Ami} S.,  {Manulis} I.,  {Gal-Yam} A.,  {Diner}
  O.,   {Rappaport} M.,  2017, in American Astronomical Society Meeting
  Abstracts \#229. p. 155.06

\bibitem[\protect\citeauthoryear{Pogosian, Tye, Wasserman  \& Wyman}{Pogosian
  et~al.}{2003}]{Pogosian:2003mz}
Pogosian L.,  Tye S. H.~H.,  Wasserman I.,   Wyman M.,  2003, \mn@doi [Phys.
  Rev.] {10.1103/PhysRevD.68.023506, 10.1103/PhysRevD.73.089904}, D68, 023506

\bibitem[\protect\citeauthoryear{Pogosian, Wyman  \& Wasserman}{Pogosian
  et~al.}{2004}]{Pogosian:2004ny}
Pogosian L.,  Wyman M.~C.,   Wasserman I.,  2004, \mn@doi [JCAP]
  {10.1088/1475-7516/2004/09/008}, 0409, 008

\bibitem[\protect\citeauthoryear{Pogosian, Wasserman  \& Wyman}{Pogosian
  et~al.}{2006}]{Pogosian:2006hg}
Pogosian L.,  Wasserman I.,   Wyman M.,  2006, {arXiv:0604141}

\bibitem[\protect\citeauthoryear{Pogosian, Tye, Wasserman  \& Wyman}{Pogosian
  et~al.}{2009}]{Pogosian:2008am}
Pogosian L.,  Tye S. H.~H.,  Wasserman I.,   Wyman M.,  2009, \mn@doi [JCAP]
  {10.1088/1475-7516/2009/02/013}, 0902, 013

\bibitem[\protect\citeauthoryear{Pshirkov \& Tuntsov}{Pshirkov \&
  Tuntsov}{2010}]{Pshirkov:2009vb}
Pshirkov M.~S.,  Tuntsov A.~V.,  2010, \mn@doi [Phys. Rev.]
  {10.1103/PhysRevD.81.083519}, D81, 083519

\bibitem[\protect\citeauthoryear{{Sako} et~al.,}{{Sako}
  et~al.}{2016}]{2016SPIE.9908E..3PS}
{Sako} S.,  et~al., 2016, in Ground-based and Airborne Instrumentation for
  Astronomy VI. p. 99083P, \mn@doi{10.1117/12.2231259}

\bibitem[\protect\citeauthoryear{Sazhin et~al.,}{Sazhin
  et~al.}{2003}]{Sazhin:2003cp}
Sazhin M.,  et~al., 2003, \mn@doi [Mon. Not. Roy. Astron. Soc.]
  {10.1046/j.1365-8711.2003.06568.x}, 343, 353

\bibitem[\protect\citeauthoryear{Sazhin, Capaccioli, Longo, Paolillo  \&
  Khovanskaya}{Sazhin et~al.}{2006}]{Sazhin:2006fe}
Sazhin M.~V.,  Capaccioli M.,  Longo G.,  Paolillo M.,   Khovanskaya O.~S.,
  2006, {arXiv:0601494}

\bibitem[\protect\citeauthoryear{Scolnic et~al.}{Scolnic
  et~al.}{2018}]{Scolnic:2017caz}
Scolnic D.~M.,  et~al., 2018, \mn@doi [Astrophys. J.]
  {10.3847/1538-4357/aab9bb}, 859, 101

\bibitem[\protect\citeauthoryear{Seljak, Slosar  \& McDonald}{Seljak
  et~al.}{2006}]{Seljak:2006bg}
Seljak U.,  Slosar A.,   McDonald P.,  2006, \mn@doi [JCAP]
  {10.1088/1475-7516/2006/10/014}, 0610, 014

\bibitem[\protect\citeauthoryear{Siemens, Creighton, Maor, Ray~Majumder, Cannon
   \& Read}{Siemens et~al.}{2006}]{Siemens:2006vk}
Siemens X.,  Creighton J.,  Maor I.,  Ray~Majumder S.,  Cannon K.,   Read J.,
  2006, \mn@doi [Phys. Rev.] {10.1103/PhysRevD.73.105001}, D73, 105001

\bibitem[\protect\citeauthoryear{Siemens, Mandic  \& Creighton}{Siemens
  et~al.}{2007}]{Siemens:2006yp}
Siemens X.,  Mandic V.,   Creighton J.,  2007, \mn@doi [Phys. Rev. Lett.]
  {10.1103/PhysRevLett.98.111101}, 98, 111101

\bibitem[\protect\citeauthoryear{Smoot et~al.}{Smoot
  et~al.}{1992}]{Smoot:1992td}
Smoot G.~F.,  et~al., 1992, \mn@doi [Astrophys. J.] {10.1086/186504}, 396, L1

\bibitem[\protect\citeauthoryear{Spergel et~al.}{Spergel
  et~al.}{2007}]{Spergel:2006hy}
Spergel D.~N.,  et~al., 2007, \mn@doi [Astrophys. J. Suppl.] {10.1086/513700},
  170, 377

\bibitem[\protect\citeauthoryear{{Stanek} \& {Garnavich}}{{Stanek} \&
  {Garnavich}}{1998}]{1998ApJ...503L.131S}
{Stanek} K.~Z.,  {Garnavich} P.~M.,  1998, \mn@doi [\apjl] {10.1086/311539},
  \href {http://adsabs.harvard.edu/abs/1998ApJ...503L.131S} {503, L131}

\bibitem[\protect\citeauthoryear{Stanishev et~al.}{Stanishev
  et~al.}{2018}]{Stanishev:2015eva}
Stanishev V.,  et~al., 2018, \mn@doi [Astron. Astrophys.]
  {10.1051/0004-6361/201732357}, 615, A45

\bibitem[\protect\citeauthoryear{Suhhonenko \& Gramann}{Suhhonenko \&
  Gramann}{2003}]{Suhhonenko:2002wb}
Suhhonenko I.,  Gramann M.,  2003, \mn@doi [Mon. Not. Roy. Astron. Soc.]
  {10.1046/j.1365-8711.2003.06175.x}, 339, 271

\bibitem[\protect\citeauthoryear{Tuntsov \& Pshirkov}{Tuntsov \&
  Pshirkov}{2010}]{Tuntsov:2010fu}
Tuntsov A.~V.,  Pshirkov M.~S.,  2010, \mn@doi [Phys. Rev.]
  {10.1103/PhysRevD.81.063523}, D81, 063523

\bibitem[\protect\citeauthoryear{Tye, Wasserman  \& Wyman}{Tye
  et~al.}{2005}]{Tye:2005fn}
Tye S. H.~H.,  Wasserman I.,   Wyman M.,  2005, \mn@doi [Phys. Rev.]
  {10.1103/PhysRevD.71.103508, 10.1103/PhysRevD.71.129906}, D71, 103508

\bibitem[\protect\citeauthoryear{Vachaspati \& Vilenkin}{Vachaspati \&
  Vilenkin}{1985}]{Vachaspati:1984gt}
Vachaspati T.,  Vilenkin A.,  1985, \mn@doi [Phys. Rev.]
  {10.1103/PhysRevD.31.3052}, D31, 3052

\bibitem[\protect\citeauthoryear{Vilenkin}{Vilenkin}{1981}]{Vilenkin:1981zs}
Vilenkin A.,  1981, \mn@doi [Phys. Rev.] {10.1103/PhysRevD.23.852}, D23, 852

\bibitem[\protect\citeauthoryear{Vilenkin}{Vilenkin}{1984}]{Vilenkin:1984ea}
Vilenkin A.,  1984, \mn@doi [Astrophys. J.] {10.1086/184303}, 282, L51

\bibitem[\protect\citeauthoryear{Vilenkin \& Shellard}{Vilenkin \&
  Shellard}{2000}]{Vilenkin:2000jqa}
Vilenkin A.,  Shellard E. P.~S.,  2000, {Cosmic Strings and Other Topological
  Defects}.
Cambridge University Press, \url
  {http://www.cambridge.org/mw/academic/subjects/physics/theoretical-physics-and-mathematical-physics/cosmic-strings-and-other-topological-defects?format=PB}

\bibitem[\protect\citeauthoryear{Witten}{Witten}{1985}]{Witten:1985fp}
Witten E.,  1985, \mn@doi [Phys. Lett.] {10.1016/0370-2693(85)90540-4}, 153B,
  243

\bibitem[\protect\citeauthoryear{Wyman, Pogosian  \& Wasserman}{Wyman
  et~al.}{2005}]{Wyman:2005tu}
Wyman M.,  Pogosian L.,   Wasserman I.,  2005, \mn@doi [Phys. Rev.]
  {10.1103/PhysRevD.72.023513, 10.1103/PhysRevD.73.089905}, D72, 023513

\bibitem[\protect\citeauthoryear{de Laix}{de~Laix}{1997}]{deLaix:1997dj}
de Laix A.~A.,  1997, \mn@doi [Phys. Rev.] {10.1103/PhysRevD.56.6193}, D56,
  6193

\makeatother
\end{thebibliography}
\label{lastpage}
\end{document}